%% file: mmdedisp.tex
\documentclass[iop]{emulateapj}
\newcommand{\be}{\begin{eqnarray}}
\newcommand{\ee}{\end{eqnarray}}
\newcommand{\DM}{\rm DM}

\newcommand{\DMunits}{\mbox{pc cm}^{-3}}

\newcommand{\Ibar}{\overline I}
\newcommand{\Isqbar}{\overline{ I^2}}
\newcommand{\Abar}{\overline A}

\newcommand{\Inbar} {\overline{I^n}}
\newcommand{\Insbar} {\overline{I^n_s}}
\newcommand{\Inrbar} {\overline{I^n_r}}
\newcommand{\Isbar} {\overline{I_s}}
\newcommand{\Issqbar}{\overline{I^2_s}}

\newcommand{\tDM}{t_{\rm DM}}
\newcommand{\Nnu}{N_{\rm \nu}}
\newcommand{\Nt}{N_{\rm t}}
\newcommand{\Ndm}{N_{\rm DM}}
\newcommand{\dnu}{\Delta \nu}

\newcommand{\tsamp}{\Delta t_{\rm s}}
\newcommand{\bw}{B}
\newcommand{\ddm}{\rm \Delta DM}
\newcommand{\tp}{t^{\prime}}

\newcommand{\giss}{g_{\rm DISS}}

\newcommand{\dtiss}{\Delta t_{\rm DISS}}
\newcommand{\dnuiss}{\Delta \nu_{\rm DISS}}
\newcommand{\taud}{\tau_{\rm d}}
\newcommand{\fwidth}{W_{\rm \nu}}
\newcommand{\twidth}{W_{\rm t}}
\newcommand{\mI}{m_{\rm I}}
\newcommand{\mImax}{m_{\rm I,max}}

\newcommand{\mIbb}{m_{\rm I,bb}}
\newcommand{\mIpulse}{m_{\rm I,bb}}
\newcommand{\mIspike}{m_{\rm I,s}}
\newcommand{\mIT}{m_{\rm I,T}}
\newcommand{\mA}{m_{\rm A}}

\newcommand{\SNR}{\rm SNR}
\newcommand{\SNRt}{\rm SNR_t}
\newcommand{\SNRft}{\rm SNR_{\nu t}}
\newcommand{\SNRmin}{\rm SNR_{min}}
\newcommand{\ff}{f_{\nu}}
\newcommand{\fft}{f_{\nu,t}}

\newcommand{\stda}{\sigma_{\rm A}}
\newcommand{\dtddm}{\Delta t_{\rm \delta DM}}
\newcommand{\dtp}{\Delta t_{\rm p}}

\begin{document}

\title{Multimoment Radio Transient Detection}
\author{L. G. Spitler, J. M. Cordes, S. Chatterjee}
\affil{Astronomy Department and NAIC, Cornell University, Ithaca, NY, 14853}
\and
\author{J. Stone}
\affil{Barnard College, New York, NY, 10027}
\email{lspitler@astro.cornell.edu}

%Abstract
\begin{abstract}
We present a multimoment technique for signal classification and apply it to the detection of fast radio transients in incoherently dedispersed data. Specifically, we define a spectral modulation index in terms of the fractional variation in intensity across a spectrum. A signal whose intensity is distributed evenly across the entire band has a lower modulation index than a spectrum whose intensity is localized in a single channel. We are interested in broadband pulses and use the modulation index to excise narrowband radio frequency interference (RFI) by applying a modulation index threshold above which candidate events are removed. The technique is tested both with simulations and using data from known sources of radio pulses (RRAT~J1928+15 and giant pulses from the Crab pulsar). The method is generalized to coherent dedispersion, image cubes, and astrophysical narrowband signals that are steady in time. We suggest that the modulation index, along with other statistics using higher-order moments, should be incorporated into signal detection pipelines to characterize and classify signals.
\end{abstract}

%Introduction
\section{Introduction}
\label{introduction}

\input{introduction}

%Method
\section{Method}
\label{method}
\input{mmdd}

%MERGE SECTIONS
%Implementation
%\section{Implementation}
\section{Application}
\label{implementation}
\input{implementation}
%Application
%\section{Application}
\subsection{Application to Data}
\label{application}
\input{application}

\section{Extensions of the Method}
\label{extensions}
\input{extensions}

\section{Discussion}
\label{sources}
\input{sources}

\section{Conclusion}
\label{conclusion}
\input{conclusion}

\input{mmdedisp.bbl}
\end{document}

%% file: introduction.tex
Surveys have always played an important role in astronomy and will play an increasingly important role in the future as observatories such as the Large Synoptic Sky Telescope (LSST) and the Square Kilometer Array (SKA) come online. 
The huge volumes of data generated by surveys 
require robust pipelines to identify and characterize populations 
with maximal completeness and minimal false positives. Regardless of the target source class, all detection pipelines rely on signal-to-noise ratio (SNR) to find sources and quantify their believability. This approach underutilizes the spectral information contained in the data, because it only uses the total intensity, or first moment, of a spectrum. We propose adding a second statistic, the spectral modulation index, that uses both the first moment (signal mean) and second moment (signal variance) of a spectrum to classify signals found through their SNR. 

Although the technique we present in this paper is applicable to any data collected as intensity versus time and frequency, we focus on surveys for fast radio bursts and use the modulation index to identify and remove narrowband radio frequency interference (RFI). Surveys at radio frequencies must contend with RFI because, if  not mitigated or excised, 
it can produce many false positives in processing pipelines.  
One difficulty in removing RFI is that it arises from 
a wide variety of terrestrial sources with different 
signal characteristics.   Moreover it can be episodic or simply transient in 
nature along with the astrophysical signals that we are interested in.  
RFI can be broad in time and narrow in frequency (e.g.\ Global Positioning System satellites) or broad in frequency and narrow in time (e.g.\ lightning). Some radar signals sweep in frequency and 
mimic the plasma dispersion of astrophysical bursts. 

We define fast radio bursts as having characteristic widths less than about one second, so astrophysical plasma delays, such as those
encountered in pulsar signals and in the class of transients known as
rotating radio transients \citep[RRATS,][]{mll+06}, are important. 
When applied to fast transients,
our technique builds upon the methods presented in \citet{cm03} and \citet{mc03} and that led
to the discovery of RRATs. However, the basic
idea applies to transients of any duration.   

Actual signals will blur the distinction between RFI and signals of interest
because some RFI will be broadband and some broadband astrophysical signals will show significant frequency variation. Astrophysical sources may have a spectral dependence including a simple
spectral index or stronger modulations like those seen in solar bursts. 
Compact sources of fast transients will typically show frequency modulation
from interstellar scintillation.  We consider these effects in our 
implementation of the method. 

The modulation index is not a new statistic; it has been used to measure time variations in a variety of astronomical applications, such as variability of pulsars \citep{wab+86} and extragalactic sources \citep{kjw+01} caused by interstellar scintillations. It has also been used to study properties of the solar wind \citep{ss04}.

We lay out the mathematical groundwork for multimoment dedispersion and the calculation of the modulation index in \S~\ref{method}. In \S~\ref{implementation} we discuss the implementation of the modulation index in a fast transient detection pipeline and present the results of a simulated single pulse detection pipeline. Also in \S~\ref{implementation} we apply the technique to two known sources of single pulses: Crab giant pulses and RRAT~J1928+15. The method is extended to other types of data sets in \S~\ref{extensions}. We discuss how the spectral modulation index can be used to classify bursts from a variety of astrophysical sources in \S~\ref{sources} and make concluding remarks in \S~\ref{conclusion}.

%% file: mmdd.tex
For specificity we consider broadband astrophysical signals that are
sampled as dynamic spectra; that is,  a sequence of spectra
separated in time by $\tsamp$ with $\Nnu$ frequency channels spanning a
total bandwidth $\bw$.  Most of the cases we discuss in this paper will have time-bandwidth products well in excess of unity, i.e., $\Delta t_s B / N_{\nu}\gg 1$, as is consistent
with fast-dump spectrometers used in surveys for pulsars and radio transients.
Our discussion will also concentrate on incoherent (post-detection) 
dedispersion, although we briefly discuss applications where coherent 
dedispersion is used. To illustrate the basic method, we 
consider only a simple sum over frequency to yield an intensity time series;
later we will consider interstellar dispersion delays in the sum and 
usage of the dispersion effect in discriminating astrophysical signals from
RFI.

We define the modulation index $m_I$ as the normalized standard deviation of the intensity $I$ across frequency ($\nu$),
\be
m_I^2 = \frac{\Isqbar - \Ibar^2}{\Ibar^2},
\label{eq:mi_simple}
\ee
where the first and second moments, 
$\Ibar$ and $\Isqbar$, 
respectively, 
are given by
\be
\Inbar = \Nnu^{-1} \sum_{\nu} \rm I^n(\nu).
\label{eq:first_mom}
\ee

The modulation index, $m_I$, characterizes the distribution of signal power in spectrum. If the power is distributed evenly across the band, the variance is small, and the spectrum has a low modulation index. 
A broadband pulse with a flat spectrum and infinite signal-to-noise 
ratio (SNR) has $m_I = 0$.  In the opposite extreme where the power is localized to single spectral
channel, the modulation index is $m_I \to \sqrt{\Nnu-1}$ with
increasing SNR. These simple examples illustrate how broadband astrophysical transients can be
discriminated from RFI, which often consists of narrow spikes in frequency
accompanied by time variability, by requiring that the modulation index
be less than some ceiling, $\mImax$.     

\subsection{Multimoment Dedispersion}
\label{mmdd}
The modulation index improves upon current detection schemes by including more information about the signal through the calculation of both the first and second moments. The signal processing algorithms used in the detection of short-duration radio transients (i.e., dedispersion and smoothing) must therefore be expanded to higher order moments. Although we only use the first and second moments in this paper, one could consider statistics that use higher order moments (e.g., skewness and kurtosis), so we introduce general, $n^{th}$-order expressions. 

Radio pulses traveling through the interstellar medium are subject to frequency-dependent dispersion. Standard pulsar processing techniques remove the effects of dispersion by shifting the intensity in each frequency channel in time according to the $\nu^{-2}$ dispersion relation and averaging to increase the signal-to-noise ratio (SNR). 

The standard incoherent or post-detection approach for calculating a dedispersed time series is 
\be
\Ibar(t, \DM) = \frac{1}{\Nnu} \sum_{\nu} I(t + \tDM(\nu), \nu)
\label{eq:std_dd}
\ee
where $I(t,\nu)$ is the time-frequency intensity data, $\Nnu$ is the number of frequency channels, and $t_{DM}(\nu)$ is the delay at frequency $\nu$ for dispersion measure DM. Throughout we will denote an average in frequency with a bar over the variable. 
Survey data must be dedispersed using a set of trial DMs because the DM of the target sources are not known \emph{a priori}, except in special applications, such as searches of globular clusters with previously known pulsars. A list of candidate pulses is defined by applying a minimum SNR threshold ($\SNRmin$) to the set of dedispersed time series. 

Equation~\ref{eq:std_dd} implicitly weights all frequency channels equally. 
In general though there will be effects, both astrophysical and instrumental, for which optimal detection (i.e., maximum SNR) requires unequal weighting. For example broadband astrophysical sources have spectra with slopes characterized by a spectral index $\alpha$, and maximum SNR occurs when the frequency channels are given some weighting $w_{\nu}$ that reflects $\alpha$. Similarly removing instrumental effects, like bandpass subtraction, may introduce channel-dependent root-mean-square noise. The generalized expression for a weighted, dedispersed, first moment time series is
\be
\Ibar(t, \DM; \alpha) = 
        \frac
                {\displaystyle\sum_{\nu} w_{\nu} 
			\left(\nu/\nu_0\right)^{\alpha}
			I(t + \tDM(\nu), \nu)}
                {\displaystyle\sum_{\nu} w_{\nu}}.
\label{eq:dd_freqw}
\ee
A full discussion of frequency weights is presented in Section~\ref{freq_weights}.

Optimization in detection can also be made by considering the duration of the pulse relative to the time resolution of the data. The effective time resolution can be decreased by smoothing the data, and the maximum SNR occurs when the effective time resolution of the smoothed data matches that of the pulse \citep{cm03}. The simplest method of smoothing adds adjacent time samples until all of the signal's power is in a single sample. In general any smoothing technique can be defined by applying smoothing weights $w_t$ to the time-frequency data and averaging in time;
\be
I_s(t, \nu) = \frac{\displaystyle\sum_{t'} w_{tt'} \rm I(t', \nu)} {\displaystyle\sum_{t'} w_{tt'}},
\label{eq:I_sm}
\ee
where $I_s(t, \nu)$ is the smoothed time-frequency data and $I(\tp, \nu)$ is the original time-frequency data.  A dedispersed, smoothed time series $\Isbar(t, DM)$ is calculated according to Equation~\ref{eq:std_dd} substituting $I_s$ for $I$.

%\be
%\Isbar(t, DM)=\frac{1}{N_{\nu}} \sum_{\nu} I_s(t-t_{DM}(\nu, \nu)).
%\label{eq:I_smbar}
%\ee

The calculation of the modulation index requires the second moment of the intensity time series. We make one final generalization by expanding Equations~\ref{eq:dd_freqw} and~\ref{eq:I_sm} to higher order moments. The weighted, dedispersed $n^{th}$-moment time series is
\be
\Inbar(t, \DM; \alpha) = 
        \frac
                {\displaystyle\sum_{\nu} w_{\nu} 
			\left[ \left(\nu/\nu_0\right)^{\alpha}
			I(t + \tDM(\nu), \nu) \right]^n}
                {\displaystyle\sum_{\nu} w_{\nu}}.
\label{eq:dd_nfreqw}
\ee
The smoothed, dedispersed $n^{th}$-moment time series is
\be
\Insbar(t, DM)=  \frac{1}{N_{\nu}} \sum_{\nu} \left[ \frac{\displaystyle\sum_{t'} w_{tt'} \rm I(t' + \tDM(\nu), \nu)} {\displaystyle\sum_{t'} w_{tt'}}  \right]^{\rm n}.
\label{eq:I_dd_nsm}
\ee
Because the spectral modulation index is a measure of the frequency structure, and not the temporal structure, the data are smoothed first in time, thereby consolidating the signal into a single spectrum, before it is squared and averaged.
The modulation index for smoothed data is then
\be
m_I^2 = \frac{\Issqbar - \Isbar^2}{\Isbar^2}.
\label{eq:mi_basic}
\ee

We have specified separate expressions for the dedispersed intensity with frequency weights and smoothing weights solely for clarity; an optimal detection scheme would employ both. In the remainder of the paper we will explore the role of smoothing on the modulation index but keep the assumptions that $w_{\nu} = 1$ and $\alpha = 0$. 

An alternative formulation to calculating the smoothed, dedispersed time series is to square the data before smoothing:
\be
\Inrbar(t, DM)=  \frac{1}{N_{\nu}} \sum_{\nu} \frac{\displaystyle\sum_{t'} w_{tt'} \rm I^n(t' + \tDM(\nu), \nu)} {\displaystyle\sum_{t'} w_{tt'}}.
\label{eq:ibar_r}
\ee
This approach captures both the time and frequency variation of the data encompassed by the sumations. Throughout this paper we will focus on the spectral modulation index but will discuss this ``time resolved" modulation index in Section~\ref{delta_dm} when we discuss the signature of incorrect dedispersion.

%FREQUENCY WEIGHTS
\subsection{Frequency weights}
\label{freq_weights}

As discussed in the previous section, weighting the frequency channels non-uniformly when calculating a dedispersed time series can improve the SNR of the detection. In general an astrophysical signal will have a power law spectrum 
\be
 P_{\nu t} = P_{\nu_o t} \left ( \frac{\nu}{\nu_o} \right)^{-\alpha_o}
 \ee
 where $ P_{\nu t}$ is the pulse flux density at time $t$ and channel $\nu$, $\nu_o$ is a reference frequency, and $\alpha_o$ is the source's spectral index. Frequency weights that yield the best SNR act to flatten the spectrum
 \be
 w_{\nu, \alpha} = \left ( \frac{\nu}{\nu_o} \right)^{\alpha}
 \ee
where optimal detection occurs for $\alpha = \alpha_o$. Uniform spectral index frequency weights implicitly assumes a flat spectral index of $\alpha = 0$. Pulsars have typical spectral indices of $\alpha_o = 1.6$ with significant variation ($\alpha_{o,min} = 0$  and $\alpha_{o,max} = 3$) \citep{lylg95}. Processing with an implied spectral index of $\alpha = 0$ is only ideal for the minority of pulsars with the lowest observed spectral indices. Instead, applying frequency weights corresponding to the mean pulsar spectral index would increase the SNR with little additional computational cost. 

Surveys for new classes of sources where $\alpha$ is unknown or surveys searching for extremely weak examples of a known population may warrant a search over spectral index. Dedispersing data using a set of trial spectral indices would increase the required computation by a factor equal to the number of trial spectral indices. Such a search may only be practical for offline post-processing. 

Frequency weights can also reflect frequency-dependent noise variations caused by instrumental effects. In general we can define a signal model 
$$
I_{\nu t} = b_{\nu} ( T_{\nu t} + g_{\nu t} P_{\nu t}),
$$
where $b_{\nu}$ is the bandpass shape, $T_{\nu t}$ is the system temperature,
$g_{\nu t}$ is the gain (e.g., K~Jy$^{-1}$) and $P_{\nu t}$ is the pulsar
or transient flux density. The quantity we are interested in is $P_{\nu t}$, but the quantity we measure is $I_{\nu t}$. Isolating $P_{\nu}$ requires removing the three frequency-dependent instrumental effects, which may introduce frequency-dependent rms noise. For example flattening the bandpass ($b_{\nu}$) will result in higher noise at the band edges. For systems operating at low frequencies ($\sim$ 100 MHz) and large total bandwidths, the system temperature may vary across the band due to the strong frequency dependence of the sky brightness temperature. Finally gain variations will be both time and frequency dependent due to the particularities of the instrument. Again one chooses $w_{\nu}$ such that SNR is maximized, so if the additive noise can be modeled as Gaussian white noise, weighting the channels by the inverse of their variance (i.e., $w_{\nu} \propto 1/\sigma_{\nu}^2$) results in the maximum SNR.

To estimate the importance of considering frequency-dependent noise variations, we adopted a simple model for system temperature $T_{\nu t} = 20 \rm K + 10 \rm K (\nu/\nu_o)^{-2.7}$ and a signal with a non-zero spectral spectral index. The resulting SNR after correcting for the spectral index and averaging over frequency was compared to the standard case that assumes a flat spectral index. At higher observing frequencies ($\sim$ 1 GHz) the improvement is small ($\sim 0.1 \%$) for small or large bandwidths because of the low sky brightness temperature. At lower observing frequencies ($\sim$ 100 MHz), there is an improvement in SNR by as much as $\sim$ 80\% for wide bandwidths. This is particularly relevant for observatories like the Low Frequency Array \citep[LOFAR,][]{vgn09} and Murchison Widefield Array \citep[MWA,][]{lcm+09}.

A non-zero spectral index will increase the variance of a spectrum and therefore also increase the modulation index. This is undesirable because it would mistakenly imply a lower filling factor and may result in a true signal being flagged as RFI. The magnitude of the increase depends on the observation frequency ($\nu_o$), bandwidth ($\bw$), and spectral index. For $\bw / \nu_o \sim 0.1$ the modulation index increase is of order $\sim 0.1$ for $\alpha \sim 3$. In the extreme case of $\bw / \nu_o \sim 1$, the modulation increase is of order $\sim 1$ for $\alpha \sim 3$. Current instruments have $\bw / \nu_o \sim 0.1$, so the increase in modulation index due to a source's spectral index is negligible. The trend is for new instruments to have larger bandwidths, so eventually the modulation increase will be significant enough that correcting for the spectral index becomes necessary. Furthermore in surveys that employ a spectral index search, the modulation index aids the analysis, because the trial spectral index closest to the true spectral index has the lowest modulation index. 

%MODULATION INDEX
\subsection{Modulation Index}
\label{modindex}

The modulation index is a quantitative measure of the patchiness, or modulation, of a spectrum. It differentiates signals whose power is distributed evenly across the band from those whose power is isolated to a few frequency channels. Broadband and narrowband signals have small and large modulation indices respectively. The level of modulation is parametrized by a frequency filling factor $\ff = \fwidth/\Nnu$ where $\fwidth$ is the number of channels in the spectrum that contains signal. Broadband astrophysical signals have a high filling factor ($\ff = 1$). RFI can be both narrowband or broadband with a filling factor ranging from $\ff = 1/\Nnu$ to 1. As this paper focuses on detecting broadband signals, the ultimate goal is to define a modulation index cutoff, $\mImax$, that will enable us to flag signals with low filling factors and vastly reduce the number of candidates created by a signal detection pipeline.

The analysis below assumes that the data have been searched for candidate signals by requiring a candidate to have an $\SNRt$ larger than a minimum SNR ($\SNRmin$), and the modulation index is only calculated for these candidate signals. Because we assume that our data has zero mean, either through construction in the case of simulated data or through bandpass subtraction in the case of real data, this assumption assures that $\Ibar > 0$ and $\mI$ is well-defined. Furthermore, our interpretation of $\mI$ assumes $\SNRmin > 1$, which is reasonable assumption since such a low $\SNRmin$ would result in a deluge of events. 

To derive a simple, analytical expression for the dependence of $\mI$ on $\ff$, we consider a spectrum with $\Nnu$ channels that contains two components: noise and signal. The noise is assumed to have a mean of zero and variance $\sigma_x^2$. The signal fills $\fwidth$ channels with intensity $A_i$ in channel $i$. While each $A_i$ may be different, it will be useful to define $\Abar$, the average signal intensity over $\fwidth$ channels. The average signal intensity over the entire band is $\ff \Abar$. As we are usually more interested in signal-to-noise ratios (SNR), we define a single-channel SNR, $\SNRft = \Abar / \sigma_x$, and time series SNR, $\SNRt = \sqrt{ \Nnu } \ff \SNRft$. The latter expression assumes that the standard deviation of the noise in the times series is $\sigma_x / \sqrt{\Nnu}$. 

The modulation index for this simple signal model is 
\be
\mI^2 = \frac{\Nnu}{\SNRt^2} + \frac{\mA^2}{\ff} + \frac{1-\ff}{\ff}
\label{eq:mi_param}
\ee
where we have introduced a separate signal modulation index, $\mA = \stda  / \Abar$. This modulation index characterizes the inherent frequency structure of a signal, and for this initial discussion we consider only signals with negligible structure ($\mA \ll 1$). In Sections~\ref{diss} and \ref{selfnoise} we discuss the effects of interstellar scintillation and pulsar self-noise and will introduce a non-zero $m_A$. 

To explore the behavior of the modulation index as a function of $\SNRt$ and $\ff$, we will look at the extrema of $\ff$: $\ff = 1$ and $\ff = 1/\Nnu$. It is important to note that Equation~\ref{eq:mi_param}, as well as the limiting expressions defined below, represents the ensemble average values of the modulation index. Statistically they are the average value expected for a given combinations of $\SNRt$ and $\ff$, but noise in the data will cause variation in the actual values. The expressions below are also idealized in so far as we have set $\mA = 0$.  

\begin{figure}
%\epsscale{1} 
\begin{center}
\includegraphics[scale=0.45]{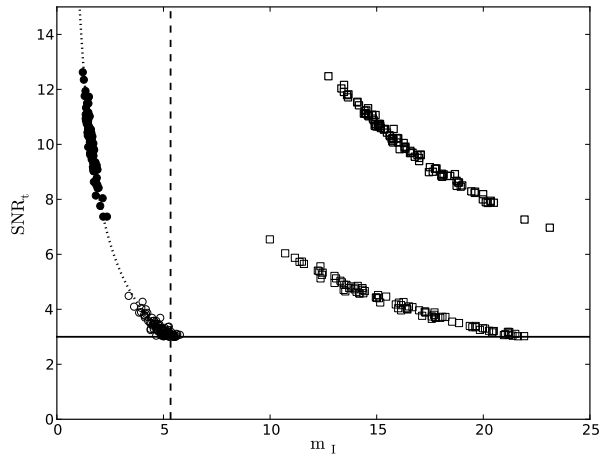}
\end{center}
\caption{Time series SNR vs. modulation index for a simulated dataset with $\Nnu = 256$ containing noise (open circles), dispersed pulses (filled circles), and two sets of one-channel-wide RFI spikes (open squares). The solid horizontal line represents the applied intensity threshold of $\SNRmin = 3$. The vertical dashed line represents $\mIT = 5.3$ as calculated for $\SNRt=3$ using Equation~\ref{eq:mi_T}. The dotted curve shows $\mIpulse$ as a function of $\SNRt$ (Equation~\ref{eq:mi_pulse}).
}
\label{simplenoise}
\end{figure}

We also simulated a time-frequency data set containing broadband and ultra-narrowband signals added to Gaussian noise to accompany the discussion. It was processed according to standard pulsar processing techniques, and the results are shown in Figure~\ref{simplenoise}.
The data set has $\Nnu = 256$ and includes 100 broadband pulses ($\ff = 1$) with $\twidth = 1$ and $\SNRt = 10$, two sets of 100 ultra-narrowband spikes ($\ff = 1/\Nnu$) with $\twidth = 1$ and $\SNRt = 5$ and $\SNRt = 10$ respectively, and $10^6$ noise-only spectra with zero mean and $\sigma_x^2=1$. We ignored dispersion for simplicity, and the time series was calculated using Equation~\ref{eq:std_dd} with $\tDM(\nu)=0$. An intensity threshold was applied to the resulting time series with $\SNRmin = 3$ (solid horizontal line), and the modulation index was calculated for samples above threshold. The vertical dashed line and dashed curve are explained below. 
%The vertical dashed line shows $\mIT = 3$, and the dashed curve is defined by Equation~\ref{eq:mi_pulse}. These will be explained below. 

A broadband signal ($\ff = 1$) with no inherent structure does not increase a spectrum's variance, so the modulation index depends only on the number of frequency channels and the signal's $\SNRt$
\be
\mIbb = \frac{ \sqrt{\Nnu} }{\SNRt}.
\label{eq:mi_pulse}
\ee
Throughout we will see that the number of frequency channels scales the modulation index but does not change the relative magnitude (i.e., signals with larger $\ff$ have lower $\mI$). As the number of channels from a single instrument generally remains constant, it is unimportant to the interpretation of a single data set but is important when comparing data from different instruments and choosing an appropriate $\mImax$.

Equation~\ref{eq:mi_pulse} reveals a direct relationship between $\SNRt$ and $\mI$. The modulation index of a broadband pulse is not arbitrary; rather, it must fall along a curve proportional to $1/\SNRt$. Furthermore for a signal with a given $\SNRt$, the broadband modulation index is the lowest $\mI$ the signal may have on average, as any $\ff < 1$ increases the frequency modulation of a signal and correspondingly its modulation index. In Figure~\ref{simplenoise} the broadband pulses (closed circles) cluster in a stripe centered at $\SNRt = 10$ and $\mIbb = 1.6$. While each pulse has an inherent $\SNRt = 10$, the underlying noise in the data spreads the $\SNRt$ and $\mI$ of individual realizations about the ensemble average value ($\mIbb$) and along the curve given by Equation~\ref{eq:mi_pulse} (dotted curve). 

In practice there is a special $\SNRt$: $\SNRmin$, the minimum SNR constraint applied to the time series in the ``thresholding" operation. The corresponding modulation index is given by Equation~\ref{eq:mi_pulse}
\be
\mIT = \frac{ \sqrt{\Nnu} }{\SNRmin}.
\label{eq:mi_T}
\ee
This modulation index represents the average maximum $\mIbb$ that may exist in a set of thresholded candidates. In real data there will be spectra with SNR above the $\SNRmin$ due to noise alone. The modulation index of thresholded noise follows the same relation as broadband pulses but with $\SNRt \sim \SNRmin$. Most of the events due to noise will cluster near $\mIT$ and $\SNRmin$, but the rare stronger noise events will pepper the broadband curve toward larger $\SNRt$ and lower $m_I$. In Figure~\ref{simplenoise} thresholded noise is shown as open circles, and only 100 points are plotted to reduce clutter. The weakest noise clusters near $\SNRmin = 3$ and $\mIT = 5.3$, and the stronger noise climbs the $\mI \propto 1/\SNRt$ curve (dotted curve).

Because $\mIT$ is on average the largest value of modulation index for a broadband signal in a candidate list, it is an upper limit to the choice of $\mImax$. Choosing a $\mImax > \mIT$ would only return events from thresholded noise or RFI. The dashed vertical line in Figure~\ref{simplenoise} shows $\mIT$ for $\SNRmin = 3$. If a modulation index upper limit is applied at $\mImax = \mIT$, events to the left of the line are kept (all of the broadband pulses and about 85\% of the thresholded noise), while all events to the right are dropped (all the ultra-narrowband spikes and about 15\% of noise events). A larger $\SNRmin$ would raise the solid line and move the dotted line to the left, which reduces the false alarm rate due to noise but limits one to detecting stronger pulses that are presumably rarer. 

For the ultra-narrowband case where $\ff = 1/\Nnu$ and $\SNRt \rightarrow \infty$, the dependence on $\SNRt$ drops out, and the average modulation reduces to 
\be
\mIspike = \sqrt{\Nnu -1}. 
\label{eq:mi_spike}
\ee
Note that the modulation index for spiky signals depends only on the number of channels. Equation~\ref{eq:mi_spike} also gives the upper limit on $m_I$ as any $\ff > 1/\Nnu$ reduces the modulation and decreases the modulation index. Spiky RFI is illustrated in Figure~\ref{simplenoise} as  two stripes of open squares centered at $\SNRt = 10$ and 5 respectively and $\mIspike = 16$. Like the broadband signals, the noise spreads the points about the ensemble average value, but clearly $\mIspike$ does not depend on $\SNRt$ for narrowband signals. 

\begin{figure}
\begin{center}
\includegraphics[scale=0.45]{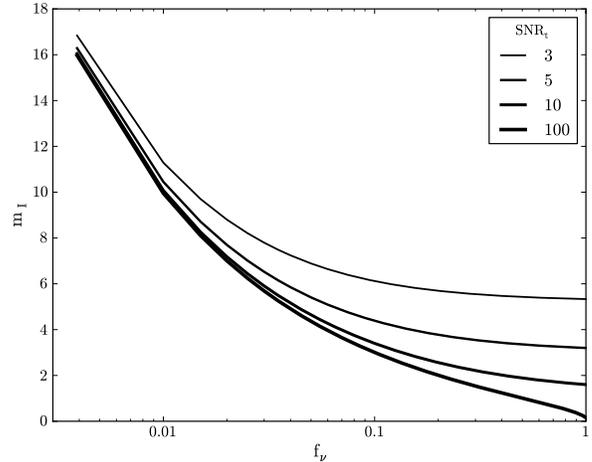}
\end{center}
\caption{Modulation index vs. filling factor calculated using Equation~\ref{eq:mi_param} with $\Nnu = 256$. From top to bottom (thin to thick), curves are shown for four values of $\SNRt$: 3, 5, 10, 100. 
}
\label{mi_snr_t}
\end{figure}

The modulation index for intermediate filling factors at constant $\SNRt$ must transition smoothly from $\mIbb$ to $\mIspike$ as $\ff $ goes from 1 to $1/\Nnu$. The exact manner of the transition is given by Equation~\ref{eq:mi_param}. Figure~\ref{mi_snr_t} illustrates the analytical variation of modulation index with filling factor for four values of the time series SNR ranging from $\SNRt = 3$ (top) to $\SNRt=100$ (bottom) and $\Nnu=256$. At low filling factors all curves tend toward $\sqrt{\Nnu}$, and at high filling factors the stronger the signal, the lower the modulation index. Most of the drop in modulation index happens at $\ff < 0.1$, suggesting that this technique can easily classify signals with filling factors less than about 10\% but is less sensitive for signals with moderate to high filling factors. 

In the above analysis we have assumed that a signal is either narrowband or broadband. Reality is messier, and we might have overlapping signals that are both narrowband and broadband. For example a pulse might occur at the same time as persistent narrowband RFI. To explore this case we adopt a simple model looking at the modulation index of a broadband pulse with no intrinsic frequency structure contaminated by an RFI spike that is one channel wide (i.e.\ the ``ultra-narroband" case described above). Equation~\ref{eq:mi_2comp} estimates the modulation index for this two-component case:
\be
m_{\rm I,bb+s} = \frac{ \sqrt{\Nnu ( 1 + \SNR^2_{\rm t,s}) }}{\SNR_{\rm t,bb} + \SNR_{\rm t,s}},
\label{eq:mi_2comp}
\ee
where $\SNR_{\rm t,bb}$ and $\SNR_{\rm t,s}$ are the time series SNR of the broadband pulse and RFI spike respectively. Note when $\SNR_{\rm t,s} \rightarrow 0$, the above equation reduces to the expression for broadband pulses (Equation~\ref{eq:mi_pulse}), and when $\SNR_{\rm t,bb} \rightarrow 0$, the above equation reduces to the expression for ultra-narrowband spikes (Equation~\ref{eq:mi_spike}, up to the ``$-1$", which is negligible for large $\Nnu$). When $\SNR_{\rm t,bb} \sim \SNR_{\rm t,s}$, we see that $m_{\rm I,bb+s} \approx 0.5 \mIspike$. If we choose the modulation index cutoff to be $\mIT$, this implies a $\SNRmin \leq 2$ to allow $m_{\rm I,bb+s}$ to be above threshold, which is an unreasonably low threshold for most surveys. Furthermore using Equation~\ref{eq:mi_2comp}, we can estimate that in order for $m_{\rm I,bb+s} \geq \mIT$, $\SNR_{\rm t,bb}/\SNR_{\rm t,s} > \SNRmin -1$. For example if $\SNRmin = 5$, the $\SNRt$ of the pulse must be four times larger than the $\SNRt$ of the narrowband RFI. Using only the $\SNRt$ and modulation index, we could easily miss a pulse if it occurs concurrent with strong, narrowband RFI. This finding that the modulation index is more sensitive to narrowband signals than broadband signals is consistent with Figure~\ref{mi_snr_t} and suggests our method is not a substitute for RFI excision techniques that identify persistent, strong, narrowband RFI from raw data.

%INTERSTELLAR scintillations
\subsection{Interstellar scintillations}
\label{diss}
Small scale density irregularities in the ionized interstellar medium (ISM) scatter and refract radio waves. Diffractive interstellar scintillations (DISS) and refractive interstellar scintillations (RISS) are observational phenomena seen in compact radio sources due to these irregularities. Scintillations are characterized by intensity variations with typical time ($\dtiss$) and frequency scales ($\dnuiss$), and for DISS in the strong scattering regime, a time series is modulated as a random variable $\giss(t)$ with an exponential amplitude distribution and fractional intensity variations on the order of unity. More generally, DISS varies with both time and frequency ($\giss(t,\nu)$) with the diffraction timescale scaling as $\dtiss \sim \nu^{1.2}$ and bandwidth scaling as $\dnuiss \sim \nu^{4.4}$. Also note that $\dnuiss$ and the pulse broadening time ($\taud$) are Fourier transform pairs and given by $2\pi \dnuiss \taud = C_1$ \citep{cr98} where $C_1 \approx 1$. For fast transients we can safely assume that the duration of the pulse is much less than the diffraction time scale (at least for $\DM \lesssim$ a few hundreds at $\nu = $ 1 GHz), whereas the scintillation bandwidth may be of the same order as the channel resolution or bandwidth of a spectrometer.  Frequency structure caused by DISS will increase the variance and thereby increase the modulation index of a signal. 

Understanding the frequency structure of a spectrum due to strong scattering is clarified by defining three limits based on the relative sizes of $\dnu$, $\bw$, and $\dnuiss$ where $\dnu$ is the width of a single frequency channel in a spectrum \citep{cbh+04}. When the scintillation bandwidth is smaller than the channel bandwidth ($\dnuiss \ll \dnu$), the instrument effectively averages over several ``scintles" and the intensity modulations are quenched. In this case $\mA \ll 1$, and the modulation index follows the expressions in Section~\ref{modindex}. Similarly in the other extreme where there is only one scintle across the band ($0.2 \bw < \dnuiss$), the intensity variation is nearly flat over the entire bandwidth and $\mA \ll 1$. For the intermediate case where there are several distinct scintles across the band ($\dnu \lesssim \dnuiss \lesssim 0.2 \bw$), the amplitude of each scintle is exponentially distributed and $\mA \approx 1$. As instruments become increasingly wide band and $\bw / \nu \sim 1$, this situation will become ever more relevant. Examples of these three cases can be seen in \citet{cbh+04} for giant pulses from the Crab pulsar. 

Connecting this to the discussion in Section~\ref{modindex}, the resulting average modulation index of a spectrum containing Gaussian noise and a broadband signal with exponentially distributed amplitudes is given by Equation~\ref{eq:mi_param} with $\ff = 1$ and $\mA = 1$. For weak signals the leading term of Equation~\ref{eq:mi_param} dominates and the spectral modulation index increases only slightly over that for a perfectly flat signal (i.e.\ Equation~\ref{eq:mi_pulse}). For example, the modulation index of a spectrum with $\Nnu = 256$ and $\SNRmin = 5$ increases from $\mI = 3.2$ to $\mI = 3.35$ . For strong signals (i.e.\ $\SNRt \gg \sqrt{\Nnu}$) the first term in Equation~\ref{eq:mi_param} becomes negligible and $\mI \approx \mA \approx 1$ for all $\SNRt \gg \sqrt{\Nnu}$. 

%SELF NOISE
\subsection{Self Noise in the Pulsar Signal}
\label{selfnoise}
Broadband pulsar signals have been modeled as amplitude modulated noise \citep{rhc75} and as modulated, polarized shot noise \citep{c76, cbh+04}. The noise in these models corresponds
to the emission over a broad range of radio frequencies while the modulation
accounts for pulse structure.    In this context, the modulation index of 
the signal is nonzero but is still smaller than the modulation indices 
expected from RFI.   In the limit where the noise has Gaussian statistics,
the intensity modulation index of polarized noise is
\begin{equation}
m_I^2 = m_{\rm ISS}^2 + (1+m_{\rm ISS}^2)(1+d_p^2)/2,
\end{equation}
where $m_{\rm ISS}$ is the modulation index of frequency structure from
DISS and $d_p$ is the degree of polarization.   The largest
modulation is for $m_{\rm ISS} = d_p = 1$ when $m_I = \sqrt{3}$.   The observed 
modulation will be reduced if frequency structure from DISS is much broader
than the total bandwidth $B$ or if multiple pulse structures are averaged
over in a single frequency channel of the spectrometer.

%INCORRECT DEDISPERSION
\subsection{Signature of Incorrect Dedispersion}
\label{delta_dm}

A bright pulse in a survey dedispersed with a large number of trial dispersion measures will yield events at many incorrect DMs in addition to the correct one. 
The true DM will return the largest $\SNRt$ and narrowest pulse width on average. The residual pulse smearing from neighboring, incorrect DMs yields a lower $\SNRt$ and wider pulse. The larger the DM error, the smaller the SNR is on average until the SNR drops below the threshold. 
This SNR--DM signature is one of the tests that a signal is a true astrophysical pulse and not RFI. The modulation index of a pulse also increases as the DM error increases.

The two-dimensional filling factor for a dispersed pulse dedispersed with a DM error of $\delta \DM$ is
\be
\fft = \left[ 1+ \frac{\left| \dtddm \right|}{\dtp} \right]^{-1},
\label{eq:fft}
\ee
where $\left| \dtddm \right|$ is the absolute value of the residual dispersion smearing and $\dtp$ is the intrinsic width of the pulse in seconds. For a pulse that is perfectly dedispersed, $\dtddm = 0$ and $\fft = 1$, while for a large DM error, $\dtddm \rightarrow \infty$ and $\fft \rightarrow 0$. 

Ideal matched filtering of a dispersed pulse, either in the absence of dedispersion or from residual smearing from incorrect dedispersion, smooths over the duration of the pulse's dispersion sweep, consolidating the signal into a single time bin.
The signal has a lower $\SNRt$ than the pulse's intrinsic $\SNRt$ because $\fft < 1$, but the one-dimensional filling factor is still $\ff = 1$. The spectral modulation index increases slightly due  to the drop in $\SNRt$ (Equation~\ref{eq:mi_pulse}). For example, a hypothetical, unresolved Crab giant pulse detected in the data set described in Section~\ref{crabgiant} with an intrinsic $\SNRt = 30$ has $\mI = 0.75$ when dedispersed using the true DM of the Crab pulsar ($\DM=56.71\; \DMunits$) and $\mI \approx 4.3$ when dedispersed with a DM error of $\delta \DM = 2\; \DMunits$.  

The spectral modulation index does reflect the signature of incorrect dedispersion but only indirectly. A more powerful statistic would depend on $\fft$ rather than $\ff$. This is accomplished by calculating an alternative modulation index ($m_{I,r}$) using the time resolved intensity moments given by Equation~\ref{eq:ibar_r}. For the same hypothetical Crab giant pulse described above, the resolved modulation index is the same as the spectral modulation index ($\mI = 0.75$) when the pulse is correctly dedispersed, but it increases to $m_{I,r} = 24.7$ for a DM error of $\delta \DM = 2 \; \DMunits$. 

Combining the information provided by the spectral modulation index and resolved modulation index helps to identify spurious events from incorrect dedispersion. By first applying a cutoff in spectral modulation index, events are classified as either broadband or narrowband in frequency. The resolved modulation index of the broadband signals distinguish those that are also broad in time from those that are narrow in time.

%MODULATION INDEX THRESHOLD
\subsection{Modulation Index Cutoff}
\label{mi_max}

The role of the spectral modulation index in a source detection pipeline is analogous to the SNR threshold ($\SNRmin$). A signal is first classified as interesting or not based on its SNR. For candidate signals that are strong enough, the application of a spectral modulation index cutoff ($\mImax$) classifies the candidate as interesting or not based on how broadband it is. Just as $\SNRmin$ is applied to data automatically, the modulation index cutoff can be applied without human supervision. 

The choice of $\mImax$ is influenced by the characteristics of the instrumentation and pipeline, as well as the importance of astrophysical effects such as diffractive interstellar scintillations. The upper limit to $\mImax$ is given by modulation index at the SNR threshold ($\mIT$). As explained in Section~\ref{modindex}, $\mIT$ is the largest modulation index on average for a broadband pulse with a perfectly flat spectrum in a survey with an applied SNR minimum $\SNRmin$ (Equation~\ref{eq:mi_T}). Choosing $\mImax > \mIT$ only yields candidates from thresholded noise or narrowband RFI. One exception is astrophysical signals with frequency structure, such as that caused by DISS. But as we showed in Section~\ref{diss}, the increase in the modulation index for weak signals is small ($\sim$ a few percent), and one could increase $\mImax$ to allow for weak, scintillating signals with minimal increase in false positives. 

Choosing $\mImax < \mIT$ will reduce the number of false positives caused by thresholded noise and incorrectly dedispersed pulses, because it effectively applies a larger $\SNRmin$ (Equation~\ref{eq:mi_pulse}). The cost is reduced sensitivity, as on average only pulses with $\SNRt$ exceeding than the larger effective $\SNRmin$ will be below $\mImax$. But for surveys where interstellar scintillation may be important, a hard lower limit to the modulation index cutoff is set by the nature of exponential statistics; namely, $\mImax = 1$, as described in Section~\ref{diss}. 

Real signals will not always cleanly divide into broadband or narrowband. For example RFI could be marginally broadband, a real pulse could have strong, narrowband structure, or both narrowband and broadband signals could occur simultaneously. But for fast transients we also have an additional parameter: dispersion. Looking at the peak dispersion measure of a candidate signal can break the degeneracy between RFI and an astrophysical signal with the same modulation index.

\subsection{Correlation Bandwidth}
\label{ci}

While the modulation index provides information about the degree of modulation in a spectrum, it does not provide any information about the shape of the modulation. A signal with amplitude $\Abar$ localized in $N$ adjacent bins has the same variance as a signal with the same amplitude whose power is distributed in $N$ isolated bins across the band. We therefore need a new measure that reflects the distribution of a signal in frequency.

We define the characteristic bandwidth ($B_{\rm c}$) to be the typical width of the frequency structure of signal. A broadband signal has $B_{\rm c} = B$, and a narrowband signal has $B_{\rm c} \sim \dnu$. We define the fractional correlation bandwidth as a measure of the typical correlation scale compared to the total bandwidth
\be
\rm FCB = \frac{B_{\rm c}}{B}.
\label{eq:ci}
\ee 
The $B_{\rm c}$ can be estimated by calculating the correlation length from the autocorrelation (ACF) of a spectrum, which we define to be the half width half maximum (HWHM) of the first lobe of the ACF. 

\begin{figure}
\begin{center}
\includegraphics[scale=0.45]{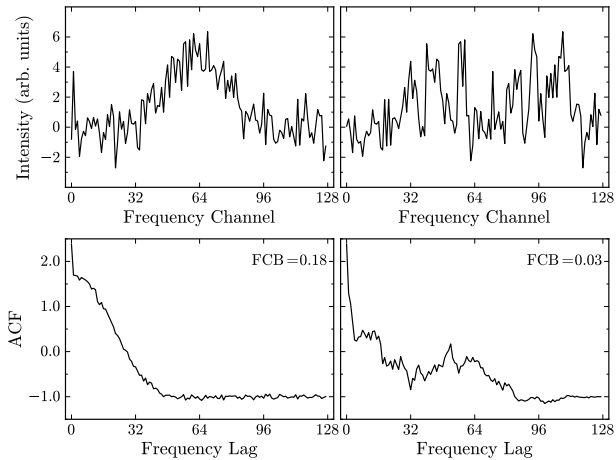}
\end{center}
\caption{Fractional correlation bandwidth (FCB) for two example spectra. The top panels show two simulated spectra with the same total intensity and variance and $\Nnu = 128$. The left spectrum contains a Gaussian with a FWHM frequency width of 32 channels, and the right spectrum is a spiky spectrum generated by randomly swapping groups of four channels in the left spectrum. The autocorrelation functions (ACF) of the sample spectra are shown in the bottom panels. The ACF has been normalized by the square of the mean of the spectrum and the number of channels all minus 1. The characteristic bandwidth is calculated as the half width at half maximum of the first lobe of the ACF, and the corresponding FCB is given in the upper right corner of the lower panels. 
}
\label{ci_plot}
\end{figure}

Figure~\ref{ci_plot} illustrates the technique for two simulated spectra with the same total intensity and variance. The left spectrum contains a Gaussian-shaped signal with a FWHM of 32 frequency channels. The right spectrum was generated by randomly swapping the channels in the left spectrum in groups of four, so that both spectra have the same total intensity, variance, and modulation index ($\mI \approx 1.5$). The bottom panels show the ACFs for the two spectra. 
To have the zero lag of the ACF equal to $\mI^2$, the ACF was scaled and offset by $\mathrm{ACF} / (\Nnu \Ibar^2) -1$, where $\Ibar$ is the mean of the spectrum. For the Gaussian spectrum, FCB = 0.18 and is consistent with the autocorrelation of a Gaussian with a FWHM=32 channels. For the spiky spectrum, FCB = 0.03 is consistent with a signal with spikes approximately four channels wide. Both spectra could be interesting astrophysical signals. The left signal could clearly be caused by an astrophysical process. The right signal could be indicative of strong interstellar scintillations. In any case the fractional correlation bandwidth provides another parameter that can be used to automatically classify a signal. Note the values for the fractional correlation bandwidth were calculated automatically along with the ACF and other statistics, suggesting the FCB could be implemented in an unsupervised pipeline.

%% file: implementation.tex
This section applies the techniques described above to the detection of fast radio transients. First we discuss one approach of incorporating the calculation of the spectral modulation index into a transient detection pipeline. To illustrate the technique we simulate a transient detection pipeline and apply the detection scheme to real data containing known transients. Our simulations and applications to real data show that a modulation index cutoff efficiently flags RFI and significantly reduces the number of candidates. Note our implementation starts with a list of pulse candidates generated in the standard manner and not on the raw data directly. A discussion of how the modulation index could be used to flag raw data in real time is discussed in Section~\ref{realtime}. Also note our technique is independent of the number of beams (e.g.\ Arecibo L-band Feed Array\footnote{\url{http://www.naic.edu/alfa}}) or stations \citep[e.g.\ Very Long Baseline Array,][]{twb+11} used to collect the data.

%\subsection{Detection pipeline}
\subsection{Implementation in a Detection Pipeline}
\label{pipeline}

The pipeline makes two passes over the data. The first pass defines a list of candidate events, and the second pass calculates the second moment and modulation index of these events. This two-pass approach is adopted for practicality and flexibility and will be explained in detail below. Also recall that the modulation index requires the first and second \emph{central} moments, so the data must be bandpass subtracted before $\Ibar$ and $\Isqbar$ are calculated. 

The data are first dedispersed as described by Equation~\ref{eq:std_dd} to generate a first moment time series. Surveys use a range of trial dispersion measures, each generating its own time series. These dedispersed time series are smoothed in time by applying a template bank of matched filters with different properties to account for pulses of different widths and profiles. A list of candidates is defined by applying a minimum SNR threshold ($\SNRmin$) to these smoothed, dedispersed time series. 
%(Note for the calculation of $\Isbar$, one can smooth in time after averaging in frequency.)
In this paper we focus on two specific implementations of matched filtering: boxcar smoothing and clustering (friends-of-friends). 

Boxcar smoothing convolves the data with a boxcar function of length $W_t$ samples. The corresponding smoothing weights for Equation~\ref{eq:I_sm} are $w_{tt'} = 1$ for $t'=0, ..., W_t$. In practice boxcar matched filtering is implemented by iteratively summing adjacent time samples so $W_t = 2^{n_{sm}}$ where $n_{sm}$ is the number of smoothing iterations. The details of this technique are described in \citet{cm03}. A single signal will likely be detected at several values of $n_{sm}$, so the event list should be sifted for the boxcar width that yields the maximum SNR.

The cluster, or friends-of-friends, algorithm looks for groupings of events in a time series. A cluster is defined as a set of events for which there is no gap in samples larger than $N_{gap}$. 
For a cluster with $N_{cluster}$ samples, the smoothing weights are $w_{tt'} = 1$ for $\tp = t_1, ..., t_N$ where $t_1$ is the first sample in a cluster, $t_N$ is the last sample, and $t'_{i+1} - t'_i < N_{gap}+1$. 
Importantly, the cluster algorithm is agnostic to the symmetry of the pulse, unlike the boxcar smoothing filter which is symmetric. As pulses from highly scattered sources have exponential tails, this algorithm may be better suited to detecting such astrophysical objects.

After a list of possible candidates is defined, one goes back to the raw time-frequency data to calculate the second moments and modulation indices. We grab a narrow range of raw data centered at the location of the candidate and reprocess it. The time-frequency snapshot is dedispersed and smoothed at the DM and smoothing parameters determined from the first pass. A time series is calculated for both the first and second moments. The first moment time series is thresholded in intensity, and the modulation index is calculated for the samples above threshold. The modulation indices of the events are compared to the modulation index cutoff, $\mImax$, and flagged as either a signal of interest if $m_I \le \mImax$ or RFI if $m_I > \mImax$. 

The two-pass approach is adopted out of practical considerations. While the first moment of the dedispersed time series can be smoothed directly, the time-frequency data must be smoothed before being squared and averaged in frequency (Equation~\ref{eq:I_dd_nsm}). This would require a different dedispersed, smoothed, second moment time series for each matched filter type and parameter. Furthermore some smoothing approaches, such as the cluster algorithm, determine the smoothing weights $w_{tt'}$ from the thresholded first moment time series, making a parallel calculation of the second moment time series impractical.

\subsection{Processing Requirements}
\label{procreq}
For most cases the two pass approach is more computationally efficient than the obvious alternative of calculating the second moment in parallel with the first. We parameterize the processing required by the second pass in terms of the processing required to do the dedispersion in the first pass. Generally dedispersion dominates the processing time in a transient survey, so this is a useful metric. The number of operations required to dedisperse a block of time-frequency data with $\Nt$ time samples and $\Nnu$ frequency channels with $\Ndm$ trial dispersion measures is $N_{\rm ops,1} = \Nt \times \Nnu \times \Ndm$. If our first pass generates $N_{\rm events}$ candidate events, the number of operations required to dedisperse a narrow time range with $N_{\rm s}$ samples around each event at a single DM is $N_{\rm ops,2} = 4 N_{\rm events} \times N_{\rm s} \times \Nnu$. The factor of 4 was included to consider the squaring of the data, summing of both the original and squared data and bandpass subtraction. The processing required by the second pass normalized by the dedispersion processing of the first pass is
\be
P_2 = \frac{N_{\rm ops,2}}{N_{\rm ops,1}} = \frac{4 N_{\rm events} N_{\rm s}}{\Ndm \Nt}.
\ee
If $\Nt = 10^6$, $\Ndm = 10^3$, $N_{\rm s} = 10^3$, $N_{\rm events} = 10^4$, $P_2$ is a few percent of the original dedispersion processing. This analysis ignores the additional overhead incurred by returning to the raw data, and in particular one must be wary of excess file I/O. 

Most RFI excision techniques operate on raw data, not on the list of candidate pulses. Our technique is more general, because it can be used at both the beginning (see Section~\ref{realtime}) and end of a source detection pipeline. The most common RFI excision approach applied to a list of candidates is to remove all events at low dispersion measure under the assumption that terrestrial signals are not dispersed. This is a blunt instrument and does not remove events from RFI at higher DMs. Calculating the modulation index of the candidate events allows for more sophisticated RFI excision in a list of candidates. 

Classifying signals with the modulation index should be used in conjunction with other RFI excision algorithms. As described in Section~\ref{modindex}, the modulation index is more sensitive to narrowband signals, and weak pulses may be missed if they occur simultaneously with strong, narrowband RFI. This suggests the modulation index algorithm works best together with algorithms that remove channels that contain persistent RFI. Similarly broadband RFI has a low modulation index, so removing impulsive RFI through other means will reduce the number of events from RFI that fall below the modulation index cutoff. 

\subsection{Simulations}
\label{simulations}

To assess the usefulness of the modulation index as a signal diagnostic, we simulated a single-pulse event detection pipeline using Python-based software. In our simulations fake time-frequency data can be generated with Gaussian-distributed noise, dispersed pulses, and a variety of RFI. A dispersed pulse is added with a specified dispersion measure and Gaussian pulse profile with a FWHM of $W_t$. Spiky RFI is modeled as a two-dimensional Gaussian with a FWHM width in time ($\twidth$) and frequency ($\fwidth$). Broadband RFI is modeled as an undispersed pulse (i.e., DM=0 $\DMunits$). The simulated data then undergo single pulse search processing. First the time-frequency data are dedispersed over a range of trial dispersion measures and both $\Ibar$ and $\Isqbar$ is calculated from the dedispersed time series. $\Ibar$ is thresholded, and for samples that are above threshold, the modulation index is calculated. Note that because these simulations involve a small amount of data, the two-pass analysis as described in Section~\ref{pipeline} is not necessary, because all intermediate data products (i.e., dedispersed time series) can be kept in memory. 

\begin{figure}
\epsscale{1} 
\begin{center}
\includegraphics[scale=0.45]{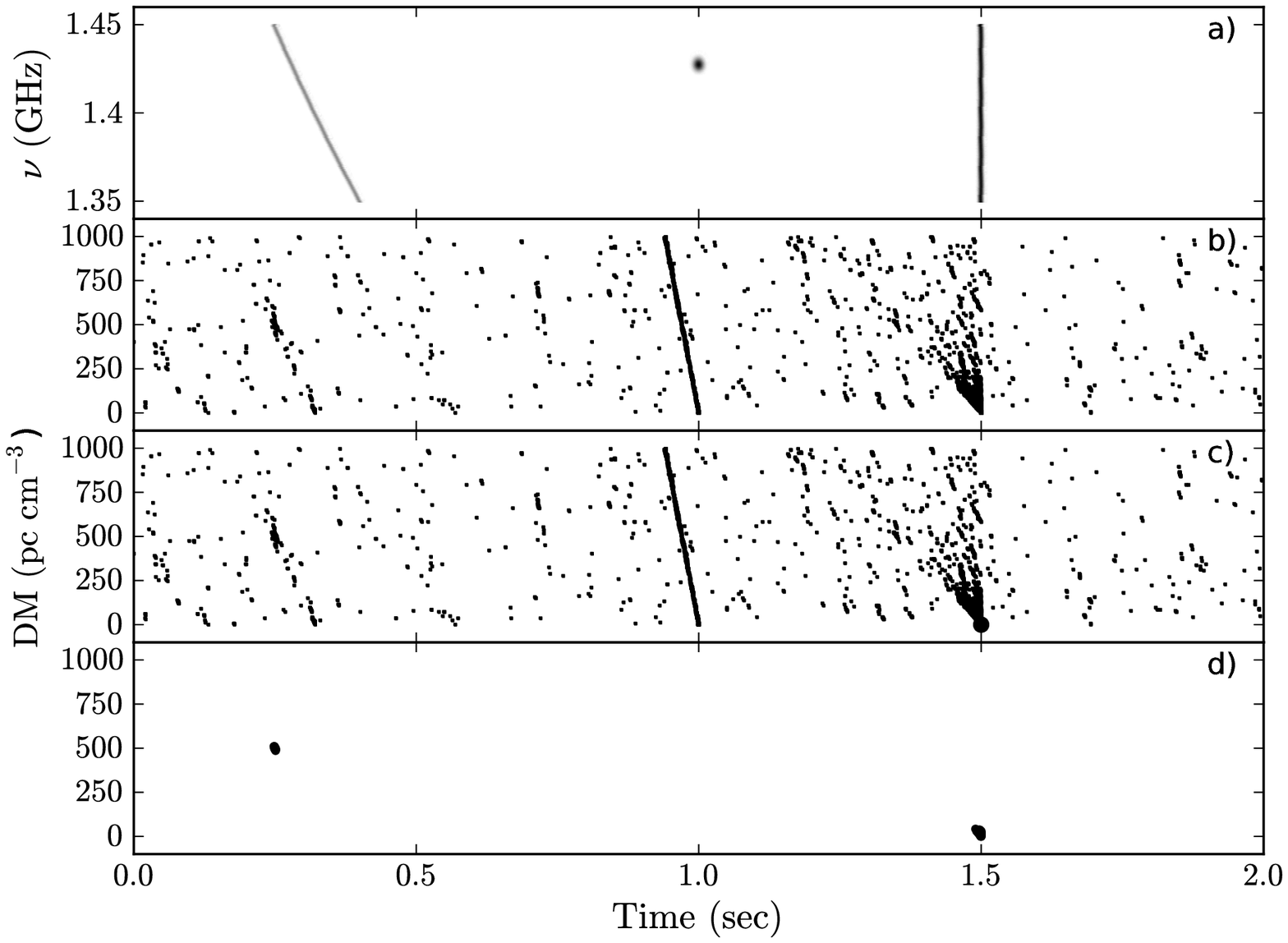}
\includegraphics[scale=0.45]{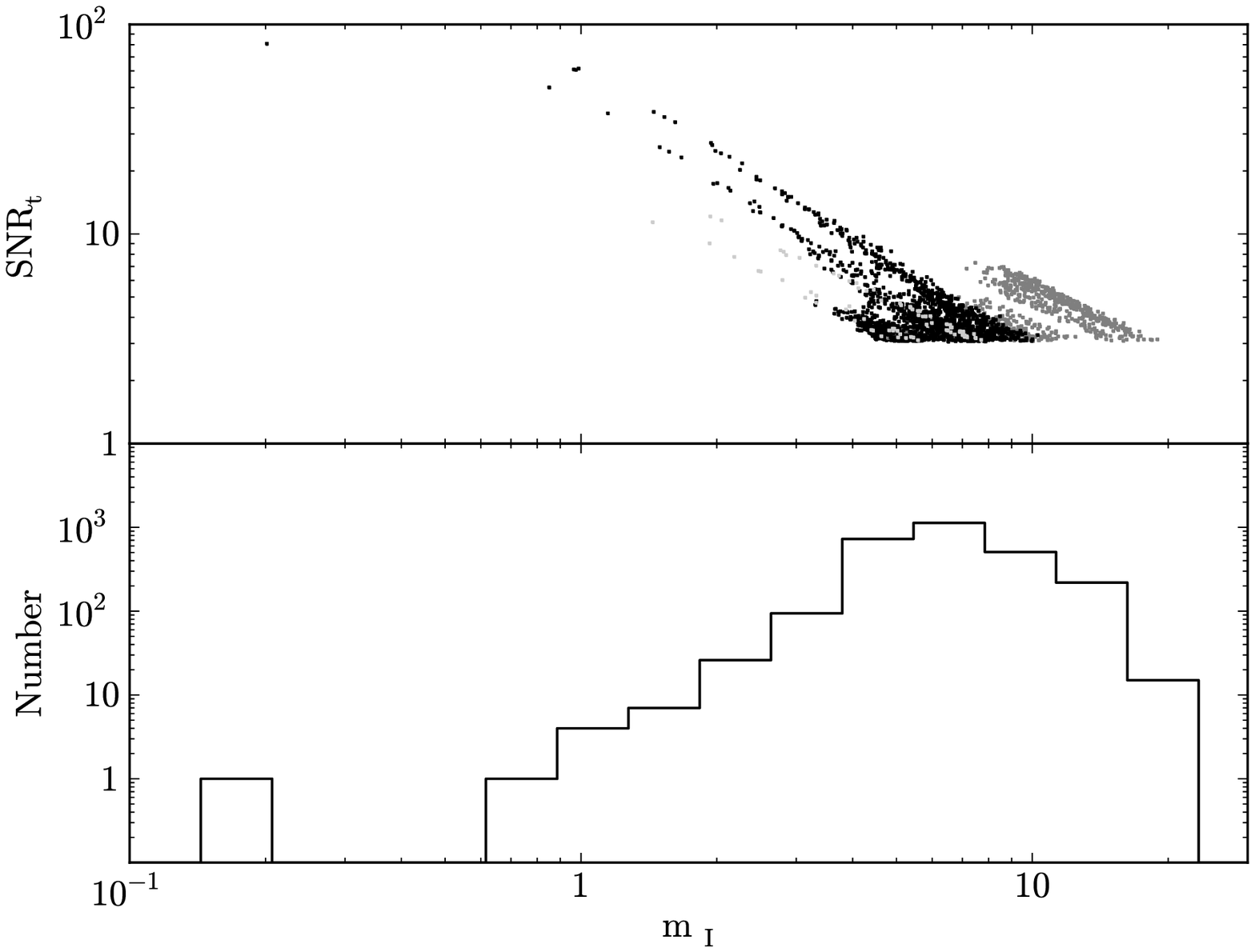}
\end{center}
\caption{Diagnostic plots for the simulated data described in Section~\ref{simulations}. Top Frame: Panel (a) shows a grey scale of the simulated time-frequency data where the SNRs and widths of the signals have been exaggerated for clarity. 
Panel (b) plots a point in the DM-time plane for each event above the intensity threshold ($\SNRmin = 3$).
Panel (c) plots $50/m_I$ for the candidate events in panel (b). 
Panel (d) plots the SNR of candidates over the SNR threshold and below the modulation index threshold ($\mImax = 3.2$). The data were dedispersed over a DM range of 0 to 1000 $\DMunits$ with $\ddm = 6 \; \DMunits$. The dispersed pulse is located at $\DM = 500\; \DMunits$ and at $t=0.5$~s. A Gaussian RFI spike is located at $t=1.0$~s and a frequency of $\sim$ 1428~MHz and causes the stripe of events between $t \approx 0.9 - 1.0$~s. Broadband RFI is located at $t=1.5$~s and causes the triangle-shaped patch of candidate events. Bottom Frame: The top panel plots $\SNRt$ vs.~$\mI$, and the bottom panel is a histogram of modulation index. Events associated with the broadband RFI (and thresholded noise) are plotted in black, events associated with the Gaussian RFI spike are shown in medium gray, and the events associated with the dispersed pulse are plotted in light gray.
}
\label{basic_dmtime}
\end{figure}

The fake data shown in panel (a) of Figure~\ref{basic_dmtime} set contains a single dispersed pulse, a Gaussian RFI spike, and a broadband RFI spike. The data properties are $\Nnu = 256$,  $\tsamp = 1$~ms, $N_{\rm t} = 2000$ time samples, $\nu_o = 1400$~MHz, and $B = 100$~MHz. A single dispersed pulse was added at $t=0.25$~s with $\DM = 500 \; \DMunits $, $\SNRft = 1$, and $\twidth = 1$. A narrowband Gaussian spike was added at $t=1.0$~s and $\nu_o \approx 1428$~MHz with $\SNRft =  40$, $\twidth = 2$, and $\fwidth = 2$. Finally a broadband RFI spike is represented by an undispersed pulse (i.e., DM = 0~$\DMunits$) at $t=1.5$~s with $\SNRft = 5$ and $\twidth = 1$. The data were dedispersed over a range of trial dispersion measures $\DM = 0 - 1000 \; \DMunits$ and DM interval $\ddm = 6 \; \DMunits$.  A list of candidate event was defined by applying an SNR threshold of $\SNRmin=3$ to the resulting time series.

Figure~\ref{basic_dmtime} shows the results of this simulation. The top frame shows (a) the time-frequency data, (b) candidate events vs. DM and time, (c) 50/$\mI$ for the candidate events, and (d) intensity vs. DM and time for candidates below $\mImax=3.2$. For clarity the SNR and pulse widths of the three signals in panel (a) have been exaggerated, and the noise is not shown. Instead of plotting the intensity, panel (b) shows a dot for each sample that was above the intensity threshold to reduce clutter.  Panel (c) shows that most of the events have similar modulation indices; the one exception is for the broadband pulse, which has a modulation index almost an order of magnitude lower than all other points due to its high SNR. Panel (d) plots the SNR of the events with modulation indices below $\mImax$ with the area of the circle proportional to the SNR of the event.

The triangle-shaped group of events near $t \sim 1.5$~s in panels (b) and (c) are spurious hits caused by the dedispersion path crossing the bright, broadband RFI samples. Most of these events have low filling factors because the RFI samples contribute only a few samples to the dedispersed spectrum and are not present in panel (d). The exception is low DM and $t=1.5$~s where signal from the RFI contributes to many frequency channels resulting in a low modulation index. This is the incorrect dedispersion effect described in Section~\ref{delta_dm}. While this example of RFI does pass our modulation index filter, the low DM of the event exposes it as RFI. Applying yet another filter that removes events at low DM will remove such events. The narrowband Gaussian spike causes a stripe of spurious events between $t \approx 0.9 - 1.0$~s as the path of the each trial dispersion measure crosses the spike. But for reasons just described above, these spurious candidates have modulation indices above our threshold. Finally the true astrophysical dispersed pulse, which is buried in the middle panels of Figure~\ref{basic_dmtime}, stands out in panel (d) at $t=0.25$~s after applying the modulation index cutoff. 

The bottom frame in Figure~\ref{basic_dmtime} plots $\SNRt$ vs.~$\mI$  and a histogram of $\mI$ to illustrate how the modulation index groups signal types. The spurious events caused by the Gaussian spike (medium gray) are clumped together on the far right of the plot between $8 < \mI < 20$, consistent with the extreme narrowband case given by Equation~\ref{eq:mi_spike} ($\mIspike = 16$). The events due to the dispersed pulse and broadband RFI follow the light gray and black tracks respectively. The tracks are caused by incorrectly dedispersing the signal, and in both cases the lowest modulation index corresponds to the correct dispersion measure. The events from thresholded noise (black) are clumped near $\mIT \sim 5$ and $\SNRt \sim \SNRmin$ as predicted by Equation~\ref{eq:mi_T}. Because the weaker events associated with the broadband RFI overlap with the area containing thresholded noise, we've plotted them with the same color. The histogram of $\mI$ in the lower panel shows that by applying a $\mImax = 3.2$, we have eliminated most of the events. 

%% file: application.tex
In the next two subsections we apply our modulation index classification technique to two known sources of single pulses detectable by single-pulse search pipelines. In Section~\ref{rrat} we apply the method to RRAT J1928+15 and in Section~\ref{crabgiant} to giant pulses from the Crab pulsar. Although in both cases we know the correct dispersion measure of the source, we re-analyze the data at a range of trial dispersion measures to recreate typical survey results. 

The data were bandpass corrected by dividing by the median bandpass and subtracting off the mean. Dividing by the median spectrum flattens the spectra, assuring the bandpass shape is not contributing to the variance calculation. We choose the median because it is less sensitive to extreme values caused by pulses or RFI. Subtracting the mean assures us that our spectra have zero mean.

%\subsection{RRATs}
\subsubsection{RRATs}
\label{rrat}

RRAT J1928+15 was discovered by \citet{dcm+09} in the PALFA survey running at the Arecibo Observatory. (See  \citealt{dcm+09} for the details of the observation and PALFA parameters.) In brief, J1928+15 was discovered using the single-pulse search algorithms implemented by the Cornell pulsar search pipeline. Three pulses were detected with an interval of 0.403~s at $\DM = 242 \; \DMunits$. We reprocessed the bandpass corrected data with 642 trial dispersion measures roughly equally spaced over a DM range $\DM = 0 - 500 \; \DMunits$. The dedispersed time series had an intensity threshold applied at $\SNRmin = 4$. The samples above the SNR threshold were run through the cluster algorithm, and the sample in each cluster with the maximum intensity was used to calculate the modulation index. 

\begin{figure}
\begin{center}
\includegraphics[scale=0.45]{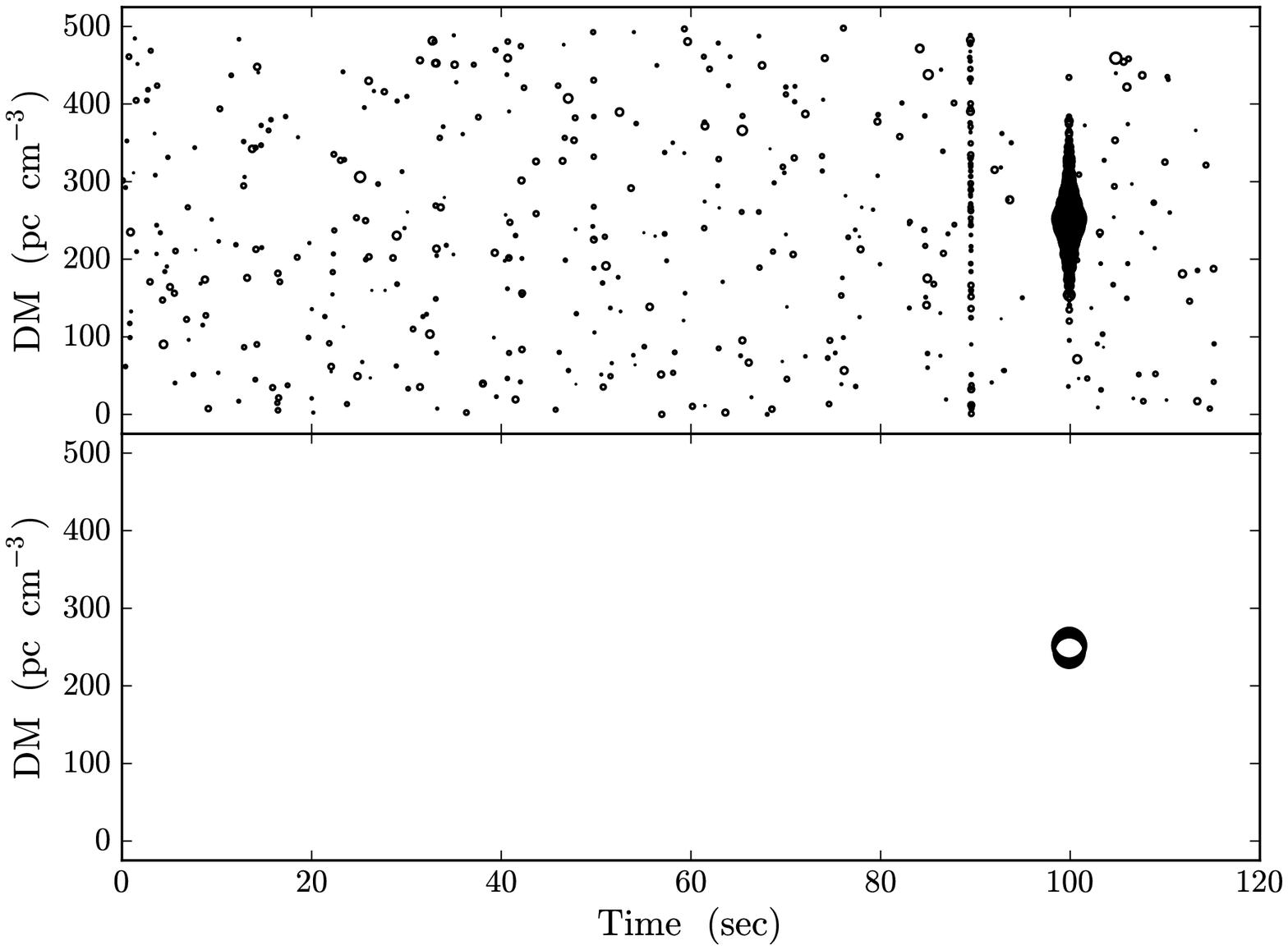}
\includegraphics[scale=0.45]{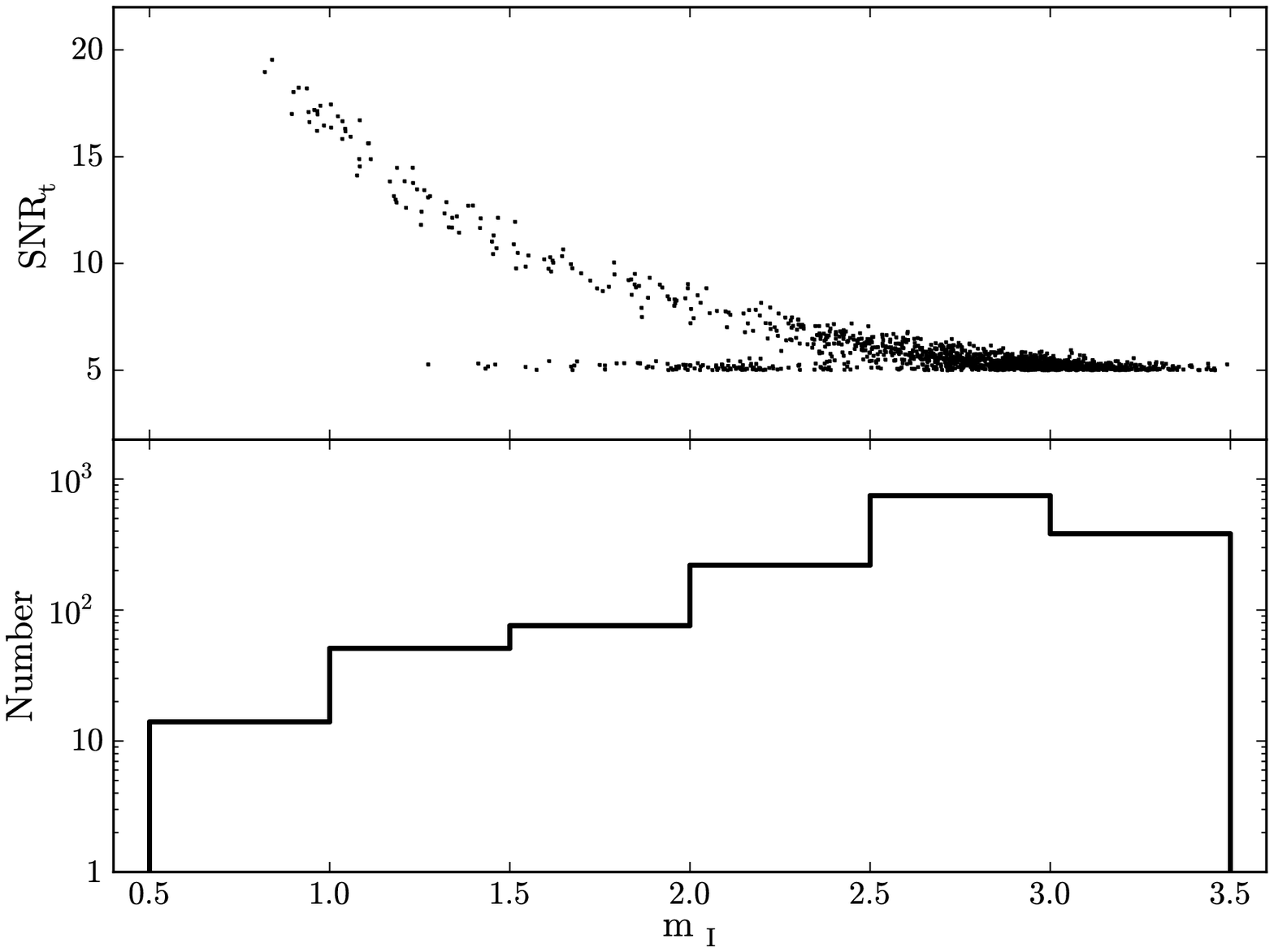}
\end{center}
\caption{Modulation index calculation for the RRAT J1928+15. In all frames events with $\SNR > 5$ are shown. Top Frame: Intensity vs. DM and time (top) for the events above the SNR threshold and the same events (bottom) with an additional modulation index threshold ($\mImax < 1$) applied (bottom). The area of each circle is proportional to the SNR of the event. (See \citealt{dcm+09}, Figure 3.) The RRAT pulse is clearly visible around $t=100$~s and DM = 242 $\DMunits$ and a weak, broadband RFI spike around $t = 89$~s. Bottom Frame: The $\SNRt$ vs. $\mI$ (top) and a histogram of $\mI$ (bottom) . The data were dedispersed using $\sim 640$ trial dispersion measures ranging from 0 to 500 $\DMunits$.  
}
\label{rrat_plots}
\end{figure}

The results are shown in Figure~\ref{rrat_plots}. In all panels only events with $\SNR > 5$ are shown. The top frame plots intensity versus DM and time for all events above threshold (top) and the events that also satisfy $\mI < 1$. (Compare to Figure 3 in \citealt{dcm+09}). The strong RRAT pulse is clearly visible around $\rm t = 100$ s, and a weak broadband RFI spike occurs near $t = 89$ s. For the strong RRAT pulse we measure $\SNRt \approx 20$, and for the preceding weaker pulse, $\SNRt \approx 6$. The difference between our values and those in \citet{dcm+09} is likely due to the fact that we used the cluster algorithm and they used the boxcar smoothing matched filter algorithm. We also do not detect the weaker, tailing pulse, likely for the same reason. Applying a $\mImax = 1$ eliminates $\approx 99\%$ of the events, leaving only the strongest events associated with the bright RRAT pulse. 

The bottom frame shows $\SNRt vs.\ m_I$ (top) and a histogram of $\mI$ (bottom). In the top panel the events associated with the strong RRAT pulse are localized along the upper curve extending from $\SNRt \approx 20$ and $\mI \approx 0.8$ to $\SNRt \approx 5$ and $\mI \approx 3$. The lower limit of $\mI \approx 0.8$ is consistent with the value predicted by Equation~\ref{eq:mi_pulse} for $\SNRt=20$. The curve itself comprises spurious events caused by the strong pulse being dedispersed at incorrect dispersion measures and account for about 90\% of all events in this figure. As described in Section~\ref{delta_dm}, applying a modulation index cutoff can reduce the number of spurious events from real signals at incorrect DMs. The broadband RFI spike produced the events along the vertical line along $\SNRt = 5$ and $1.25 < \mI < 2.5$. The events associated with thresholded noise are grouped between $2.5 < m_I < 3.5$ and are consistent with $\mIT = 3.2$ for $\Nnu = 256$ and $\SNRmin = 5$. While applying $\mImax = 1$ cleanly isolated the bright pulse from the RRAT, we did so at the cost of applying an effective $\SNRmin = 10$ and thereby removed the weaker RRAT pulse. 

%\subsection{Crab giant pulses}
\subsubsection{Crab giant pulses}
\label{crabgiant}

The most well-known source of single, dispersed pulses is the Crab pulsar. We reprocessed a 430 MHz data set containing the ``supergiant" pulse with a $\SNRt \sim 1500$ reported in \citet{cbh+04}. The 140-second long dataset was dedispersed using 616 trial dispersion measures ranging from $\DM = 0 - 250 \, \DMunits$. An intensity threshold of $\SNRmin = 4$ was applied to the dedispersed time series, and the samples above threshold were run through the cluster algorithm. The modulation index was calculated for the sample from each cluster that had the maximum intensity. This data set proved to be a particularly interesting case study due to the large number of pulses with a range of SNRs. 

\begin{figure}
\begin{center}
\includegraphics[scale=0.45]{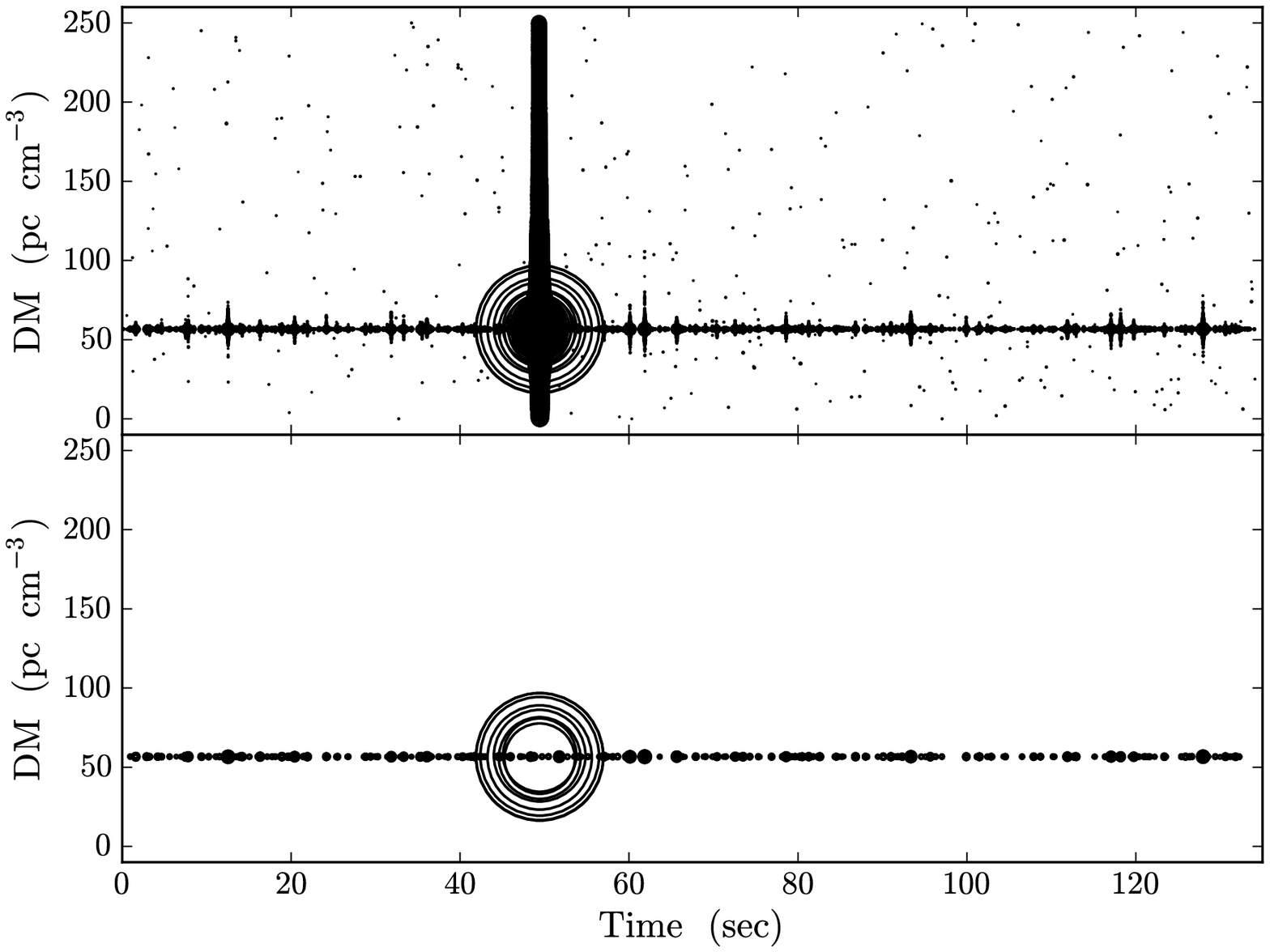}
\includegraphics[scale=0.45]{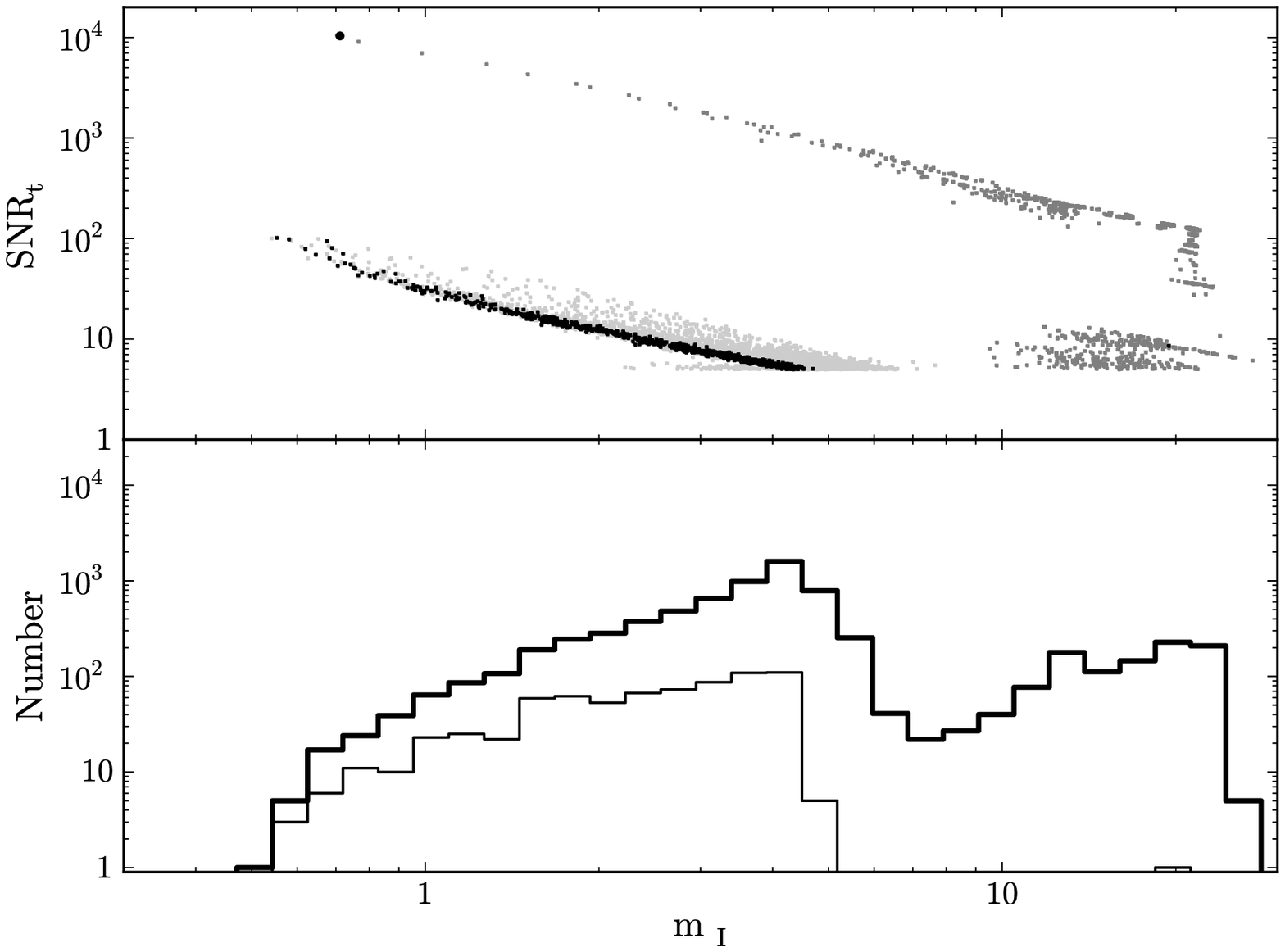}
\end{center}
\caption{Modulation index calculation for the Crab pulsar at 430 MHz. In all frames only events with $\SNR > 5$ are shown. Top Frame: Events above the intensity threshold are plotted against DM and time (top) and with the additional constraint $\mImax < 2$ (bottom). A stream of Crab giant pulses can be seen in the DM-time plot at DM=56.71 $\DMunits$. The supergiant pulse discussed in \citet{cbh+04} occurs around $t=50$~s. Bottom Frame: The $\SNRt$ vs. $m_I$ is plotted in the top panel. The points associated with the supergiant pulse are plotted in medium gray, points associated with the dispersion measure of the Crab are shown in black, and all other points shown in light gray. A $m_I$ histogram (bottom) plots the total number of events (thick line) and the number of events at DM=56.71 $\DMunits$ (thin line) as a proxy for the number of pulses detected.  
}
\label{crab_plots}
\end{figure}

The results of the reprocessing are shown in Figure~\ref{crab_plots}, and in all panels only events with $\SNR > 5$ are shown. The top frame shows the intensity for the events versus $\DM$ and time with a circle whose area is proportional to the SNR of each event. The top panel shows all events, and the bottom panel shows those events that also fall below our modulation index cutoff ($\mImax < 2$). A train of normal giant pulses is clearly visible along the dispersion measure of the Crab, $\DM = 56.71\; \DMunits$, and the supergiant pulse is located near $t = 50$~s. This pulse is so bright that it has some of the same characteristics as broadband RFI; namely, events occurring at high dispersion measures. But as we saw previously for J1928+15, applying a $\mImax$ eliminates many of the spurious events at incorrect DMs because they have low filling factors. 

The bottom frame plots $\SNRt$ vs.\ $m_I$ (top) and a histogram of $\mI$ (bottom). The events associated with the supergiant pulse are shown in medium gray and fall along the curve extending from $10^4 > \SNRt > 100$ and $0.7 < \mI < 11$ as well as the clump at $\SNRt \sim 10$ and $\mI \sim 15$. Note that the largest modulation indices are consistent with the value for spiky RFI ($\mIspike \approx 21$). The correctly dedispersed normal giant pulses are shown in black and lie along the left edge of the main cluster ranging from $100 > \SNRt > 5$ and $0.5 < \mI < 3$. One exception is the single, larger black point at the low-$\mI$ end of the supergiant pulse curve (medium gray) showing the point associated with correct dedispersion value. Not surprisingly this event has the largest $\SNRt$ and lowest $\mI$. The events from incorrectly dedispersing the regular giant pulses (light gray) spread to lower $\SNRt$ and higher $\mI$ than their correctly dedispersed counterparts (black). Finally the noise (light gray) falls between $2 \lesssim m_I \lesssim 7$ and along $\SNRt \sim 5$ and is consistent with $\mIT \approx 4.2$. 

The measured values of $\mI$ for the Crab pulses are systematically larger than that predicted by Equation~\ref{eq:mi_pulse} for broadband pulses. The brightest regular giant pulses have $\SNRt \approx 100$ and $\mI \approx 0.5$, but the modulation index of a perfectly flat pulse with this $\SNRt$ is $\sim 0.2$. This suggests that Crab giant pulses have significant inherent frequency strucutre and non-zero $\mA$. Most striking is that the modulation index for the supergiant pulse ($\mI \approx 0.7$) is larger than the modulation index for the strongest regular pulses even though its SNR is two orders of magnitude larger. The spectrum of the supergiant pulse shows significant variation across the band due to DISS, as shown in \cite{cbh+04}. 

The histogram of $\mI$ shows the total number of events at all DMs (thick line) and the number of events at the DM of the Crab (thin line) as an proxy for the number of giant pulses. The assumption that any event at the DM of the Crab is a giant pulse is simplistic and may include false positives. Because brighter giant pulses are rarer than weaker pulses, Equation~\ref{eq:mi_pulse} tells us that the number of pulses in a $\mI$ bin will decrease as $\mI \rightarrow 0$. There is therefore a tradeoff between choosing a higher $\mImax$ to allow for the more common, weaker pulses and the false alarm rate. 

\begin{figure}
\begin{center}
\includegraphics[scale=0.45]{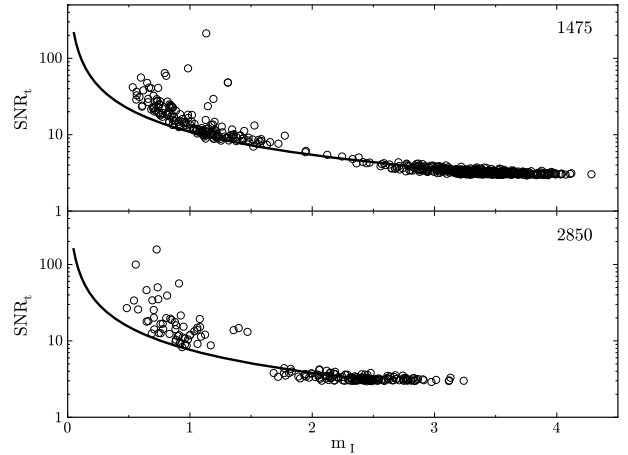}
\end{center}
\caption{$\SNRt$ vs $m_I$ of Crab giant pulses at 1475 MHz (top) and 2580 MHz (bottom) from \citet{cbh+04}. The lines represent $\mIpulse$ (Equation~\ref{eq:mi_pulse}). The modulation index that divides the pulses from thresholded noise is $\mI \approx 2$ for 1475 MHz and $\mI \approx 1.5$ for 2850 MHz. The modulation indices of the thresholded noise are consistent with those of broadband pulses. The modulation indices of the pulses are systematically larger than what is predicted by Equation~\ref{eq:mi_pulse}, suggesting the spectra have inherent structure caused by diffractive interstellar scintillations.
}
\label{crab_mf}
\end{figure}

Figure~\ref{crab_mf} further explores the role of interstellar scintillations on the modulation index. The $\SNRt$ vs. $\mI$ is plotted for two higher frequencies from \citet{cbh+04}. The data are time-frequency snapshots of pulses determined from previous processing at 1475 and 2850 MHz. The snapshots were only dedispersed at the dispersion measure of the Crab pulsar, the time series were thresholded with $\SNRmin = 3$, and the resulting candidates were run through the cluster algorithm to find the sample with the maximum SNR. Also plotted is the $\mIpulse$ curve given by Equation~\ref{eq:mi_pulse} with $\Nnu = 118$ at 1475 MHz (top) and $\Nnu = 58$ at 2850 MHz (bottom). 

In both panels there are two clusters of points: one due to thresholded noise and one due to the pulses. The thresholded noise lies below $\mI =2 $ and $\mI= 1.5$ at 1475 and 2850 MHz respectively. The typical values for the noise are consistent with $\mIT$ and lie along the broadband curve as predicted in Section~\ref{modindex}. The modulation indices of the pulses are systematically larger than what is predicted, suggesting the pulses have inherent frequency structure and $\mA \neq 0$. The spectra for the strongest pulse at each frequency are shown in \citealt[][Figures 7 and 9]{cbh+04}, and in both cases they exhibit frequency structure caused by DISS. 

%% file: extensions.tex
Our discussion has focused the detection of radio bursts using incoherent dedispersion, but the technique has more general applicability. It can also be applied to other classes of signals (periodic objects like pulsars, spectral lines) and other data formats (coherent dedispersion, image cubes from interferometers). 

\subsection{Periodic Signals}
\label{periodic}

Using the modulation index to characterize the frequency structure of a single pulse can be extended to periodic signals. The procedure is best illustrated by considering a dedispersed time series $\Ibar(t, DM)$ that has been folded at some period $P$. A pulse in this folded time series is the sum of $N_{\rm p}$ pulses 
\be
\Ibar_p(t_{\rm o}, DM) = \frac{1}{N_{\rm p}} \sum_{\rm t_p} \Ibar(t, DM)
\label{ibar_p}
\ee
where $t_{\rm p} = t_{\rm o} + n P$, $t_{\rm o}$ is the time sample of the candidate pulse in the folded time series, and $n = 0 \ldots N_{\rm p} - 1$. The modulation index is then calculated for samples where $\Ibar_p(t, DM)$ is above the SNR threshold. As discussed in Section~\ref{mmdd} for smoothing in time, calculating the second moment requires that spectra are first summed in time before averaging over frequency, making it obligatory that one returns to the time-frequency data
\be
\overline{I_p^2}(t_{\rm o}, DM) = \frac{1}{\Nnu} \sum_{\rm \nu} \left[ \frac{1}{N_{\rm p}}  \sum_{\rm t_p} I(t, \nu, DM)\right]^2.
\label{isqbar_p}
\ee
One can also calculate the second moment for the Fourier analysis method of finding pulses signals but that is beyond of the scope of this paper.

The additional computational cost required to calculate the modulation index can be estimated in the same manner as for the single pulse detections described in Section~\ref{procreq}. But as a pulsar is a repetitive source, the time span of the data that would need to be reprocessed in a second pass would be larger than for a single pulse. In the extreme case where $N_{\rm s} = \Nt$, the processing load is comparable when the number of candidates is of the order of the number of trial DMs. It is therefore prudent to reduce the number of candidates by other means first before calculating the modulation index of candidate pulsar. 

\subsection{Coherent Dedispersion}

Coherent dedispersion operates on the basedband voltage directly rather
than on intensity-like quantities at the output of a spectrometer, which our previous discussion has centered on.    In a survey, a set of trial values for DM would be used, as with post-detection dedispersion described earlier.  With the correct value of DM, coherent dedispersion restores the phase of the signal to what it was prior to alteration by the ISM (so long as multipath scattering is not important).  The procedure deconvolves a phase function from voltage data $\epsilon(t)$ by applying a
complex phase filter \citep[e.g.,][]{hr75} to produce
the dedispersed voltage $\epsilon_d(t)$.   The detected signal
$I_d(t) = \vert \epsilon_d(t)\vert^2$ would then be analyzed with
a SNR threshold to identify candidate bursts.

If the true signal is an unresolved pulse with width $W\ll B^{-1}$ where,
as above, $B$ is the total bandwidth, the dedispersed output will have
a width $W_t \approx B^{-1}$ only if the signal is unmodulated across
the band. This requires that any scintillation modulation have a characteristic bandwidth $\gg B$. In this instance, no new information is gained by analyzing  the data in the combined frequency-time plane.

For signals that have broader time extent, however, the analysis presented
in the paper for post-detection dedispersion still applies.
The dynamic spectrum would be calculated for each burst
identified in $I_d(t)$ from short time blocks of
length $\Delta t_s$ using the discrete Fourier transform (DFT)
$\tilde \epsilon(t)$ of the baseband signal.   Using an $N_{\nu}-$point
DFT, the uncertainty relation implies
$\Delta t_s B/N_{\nu} = 1$.   For burst widths $W\gtrsim \Delta t_s$,
the series of spectra that span the burst  can be used to calculate the modulation index across frequency, which is then used to classify signals as before.

\subsection{Application to Images Obtained On a Frequency-Time Grid}

The modulation index as discussed so far is calculated for a set of
intensity measurements sampled in time and frequency at a particular
value of the dispersion measure, $I(t,\nu,DM)$. For signals that are
not inherently narrow in time (i.e., not pulsed or transient in
nature), no dedispersion is required.  Instead, we simply calculate the modulation index $\mI$ as
the normalized variance of the intensity across frequency at a
particular time,
\be
m_I^2 = \frac{\Isqbar(t) - \Ibar(t)^2}{\Ibar(t)^2}.
\ee
The modulation index $\mI$ and fractional correlation bandwidth FCB
calculated in this manner can be used, for example, to characterize
radio observations of a spectral line or maser source (which show
coherent frequency structure) and discriminate them from radio
frequency interference with a more random frequency structure. We discuss the application of a time-domain modulation index to these source types in Section~\ref{sources}.

The method can be further generalized to apply to interferometric
imaging observations of continuum sources that have been acquired in
multi-channel modes. While past imaging observations with the Very Large Array (VLA), for
example, have only employed a small number of channels (e.g., $\lesssim 32$) or
even just one channel in continuum mode, observations with the Expanded Very Large Array (EVLA) or
the future SKA pathfinders such as Austrailan Square Kilometre Array (ASKAP; \citealt{jfg09}) and MeerKAT \citep{bdj+09} will typically employ many more channels.  At the EVLA, the wide bandwidth available for continuum imaging (e.g., 1--2~GHz or 2--4~GHz) requires the use of many spectral channels to avoid chromatic aberration (``bandwidth smearing").  If we were to require a maximum tolerable peak response reduction of 20\% at A-array, 1--2~GHz, that would necessitate a channel bandwidth $\Delta\nu$ such that 
\[\frac{\Delta\nu}{1.5\;{\rm GHz}} \frac{\theta}{\theta_0} < 1,\] 
where $\theta/\theta_0$ is the source offset from the phase tracking center in units of the synthesized beam.\footnote{See, e.g., the EVLA Observational Status Summary, \url{http://evlaguides.nrao.edu/}}
To maintain sensitivity at that level over just half the primary beam field of view, $\theta/\theta_0 \sim 0.5 \times d / D$ where $d$ and $D$ are the dish diameter and array size respectively. For the EVLA in A-array where $d=25$~m and $D\approx36$~km, this constraint requires as many as 720 channels for continuum imaging.

One of the challenges faced by automated
source extraction pipelines working on these data is to discriminate
between low significance detections of compact sources and random
intersections of the point spread function sidelobes caused by strong
sources in the field of view.  While the exact shape of the sidelobes
depends on the details of the array and the observation, they will
scale as $I(\theta,\nu) \propto \nu^{-1}$ at an angular distance
$\theta$ from a bright source.  Therefore, an intersection of
sidelobes from different sources will show smooth structure in
amplitude as a function of channel frequency, while a true compact
source will not.  Calculating the modulation index and FCB will allow
us to exploit the extra information in the frequency channelization of
synthesis images and add an extra discriminant that is easily
implementable in automatic pipelines.

\subsection{Realtime RFI excision}
\label{realtime}
The modulation index could be useful for systems performing RFI excision on the fly. Interest in realtime incoherent dedispersion transient searches is growing thanks to the large speed-ups achieved by graphics processing units (GPUs) \citep[e.g.][]{mks+11}. As more of the signal processing moves into hardware, realtime RFI rejection is also moving into instruments. For example \citet{d10, wbd+11} have incorporated the calculation of kurtosis into the F-stage of the DiFX software correlator, and the modulation index could be calculated in an identical manner. 

In the time-frequency domain the modulation index would be most useful for identifying broadband, impulsive RFI. A sample that had an anonymously low modulation index could be flagged or blanked by the hardware. The modulation index alone should not be used to identify narrowband RFI, as a dispersed pulse is narrowband before the dispersion is correctly accounted for. 

%% file: sources.tex
In this paper we have focused our discussion on searching for radio bursts from or
similar to those from pulsars and used the modulation index, along with SNR, as a tool for RFI excision. More generally the modulation index can be one among several statistics used to {\em characterize and classify} signals rather than to filter candidates. Furthermore there may be as-yet undiscovered source classes having different time-frequency signatures. 

Low mass stars (M-dwarfs and brown dwarfs) are frequent emitters of radio bursts \citep{jkw89, b02}. \citet{bbdd90} have shown that the dynamic spectra of dMe flare stars exhibit a rich structure in both time and frequency. The spectra of the flares they observed had both broadband and narrowband components, as well as large spectral indices ($\alpha \approx 10$), suggesting various plasma phenomena at work. A large-scale survey of flare stars could use the modulation index as one of several parameters used to automatically classify flare types. Furthermore, different plasma processes result in different levels of polarization, so calculating the modulation index of each polarization separately could further aid in classifying flares and recognizing RFI. 

The study of radio variability in the solar system is a natural extension to flare stars and is another area where the modulation index could be a tool for classifying different radio bursts. Planetary auroral radio emission (ARE) has been detected for all of the magnetic planets in our solar system \citep{z98}, and dynamic spectra of ARE show complex substructure in both time and frequency. A notable class of fast radio pulses is Jupiter's ``S-bursts". These bursts have durations of milliseconds and are the brightest of Jupiter's ARE emission \citep{z98}.  Radio discharges from lightning on the gas giant planets have durations on the order of 10 to 100 ms and fluxes easily detectable at Earth \citep{zfk+04}. In fact \citet{rrk+09} have used multi-moment techniques (total intensity and kurtosis) to look for lightning on Mars during a dust storm. By extension it is expected that extrasolar planets would have radio emission that would be variable and include bright bursts \citep{fdz99, gzs07}. A study of the modulation indices of flares from Jupiter could help identify similar flares from exo-Jupiters. 

%Flares from the magnetospheres of stars are an abundant source of slow transients ~\citep{g02}. . 
%Microquasars and X-ray binaries are also known to be sources of radio flares \citep{vss+93, wgj+95}. 

Extragalactic sources of fast radio transients might include merging neutron star--neutron star or neutron star--black hole binaries~\citep{hl01}. Short gamma-ray bursts (SGRB) are thought to arise from the merger of two neutron stars, and~\citet{pp10} suggest that GRB monitors could alert low frequency radio observatories (e.g., LOFAR) to a possible radio transient. 

Our focus has been on signals that are broad in frequency but narrow in time.
The reciprocal problem of spectral lines that are steady in time can be
handled in the same way but with the roles of frequency and time reversed. 
As an example, surveys for maser lines will most likely 
require a line amplitude that is relatively steady over time scales of days 
or less.    Interstellar scintillation may induce time variations in 
some maser sources if they are compact enough.      
SETI (search for extraterrestrial intelligence) often postulates narrowband
beacon signals that are constant in time.  Interstellar scintillation 
will certainly induce time variations owing to the compact nature of any
relevant transmitters \citep{cl91} but with a modulation index
$\sim 1$. The methods outlined here apply directly to these problems. 

%% file: conclusion.tex
We have discussed how defining detection statistics based on higher order moments can improve the success of source detection pipelines and focused on the spectral modulation index. By calculating the second moment, we are able to classify signals based not only on their strength but also on their fractional frequency variation. By applying prior information about our target sources, i.e., that they are broadband, the modulation index easily distinguishes between broadband and narrowband signals and allows us to filter a large fraction false positives due to narrowband RFI. 

These detection statistics (e.g., SNR, $\mI$) are crucial to source extraction pipelines because they can be calculated automatically. As new observatories generate more and more data, it is critical that source extraction occurs reliably with minimal human intervention.  Although we limit ourselves to two statistics in this paper, a pipeline could make use of many higher-order statistics \citep[e.g.\ kurtosis,][]{ngl+07} along with a weighted voting scheme to classify signals in more sophisticated and nuanced ways.  For example the modulation index could become one parameter used by detection pipeline based on an artificial neural network \citep{emk+10}.

We thank the reviewer for his or her useful comments. This work was supported by NSF grant AST - 1008213. L.G.S. and J.S. were supported by the NASA/New York Space Grant Consortium under grant NNX10AI94H. L.G.S. also received support from the National Astronomy and Ionosphere Center (NAIC). The Arecibo Observatory is operated by SRI International under a cooperative agreement with the National Science Foundation (AST-1100968), and in alliance with Ana G. M\'{e}ndez-Universidad Metropolitana, and the Universities Space Research Association.

%% file: mmdedisp.bbl
\begin{thebibliography}{34}
\expandafter\ifx\csname natexlab\endcsname\relax\def\natexlab#1{#1}\fi
\expandafter\ifx\csname bibnamefont\endcsname\relax
  \def\bibnamefont#1{#1}\fi
\expandafter\ifx\csname bibfnamefont\endcsname\relax
  \def\bibfnamefont#1{#1}\fi
\expandafter\ifx\csname citenamefont\endcsname\relax
  \def\citenamefont#1{#1}\fi
\expandafter\ifx\csname url\endcsname\relax
  \def\url#1{\texttt{#1}}\fi
\expandafter\ifx\csname urlprefix\endcsname\relax\def\urlprefix{URL }\fi
\providecommand{\bibinfo}[2]{#2}
\providecommand{\eprint}[2][]{\url{#2}}

\bibitem[{\citenamefont{{McLaughlin} et~al.}(2006)\citenamefont{{McLaughlin},
  {Lyne}, {Lorimer}, {Kramer}, {Faulkner}, {Manchester}, {Cordes}, {Camilo},
  {Possenti}, {Stairs} et~al.}}]{mll+06}
\bibinfo{author}{\bibfnamefont{M.~A.} \bibnamefont{{McLaughlin}}},
  \bibinfo{author}{\bibfnamefont{A.~G.} \bibnamefont{{Lyne}}},
  \bibinfo{author}{\bibfnamefont{D.~R.} \bibnamefont{{Lorimer}}},
  \bibinfo{author}{\bibfnamefont{M.}~\bibnamefont{{Kramer}}},
  \bibinfo{author}{\bibfnamefont{A.~J.} \bibnamefont{{Faulkner}}},
  \bibinfo{author}{\bibfnamefont{R.~N.} \bibnamefont{{Manchester}}},
  \bibinfo{author}{\bibfnamefont{J.~M.} \bibnamefont{{Cordes}}},
  \bibinfo{author}{\bibfnamefont{F.}~\bibnamefont{{Camilo}}},
  \bibinfo{author}{\bibfnamefont{A.}~\bibnamefont{{Possenti}}},
  \bibinfo{author}{\bibfnamefont{I.~H.} \bibnamefont{{Stairs}}},
  \bibnamefont{et~al.}, \bibinfo{journal}{\nat} \textbf{\bibinfo{volume}{439}},
  \bibinfo{pages}{817} (\bibinfo{year}{2006}), \eprint{arXiv:astro-ph/0511587}.

\bibitem[{\citenamefont{{Cordes} and {McLaughlin}}(2003)}]{cm03}
\bibinfo{author}{\bibfnamefont{J.~M.} \bibnamefont{{Cordes}}} \bibnamefont{and}
  \bibinfo{author}{\bibfnamefont{M.~A.} \bibnamefont{{McLaughlin}}},
  \bibinfo{journal}{\apj} \textbf{\bibinfo{volume}{596}}, \bibinfo{pages}{1142}
  (\bibinfo{year}{2003}), \eprint{arXiv:astro-ph/0304364}.

\bibitem[{\citenamefont{{McLaughlin} and {Cordes}}(2003)}]{mc03}
\bibinfo{author}{\bibfnamefont{M.~A.} \bibnamefont{{McLaughlin}}}
  \bibnamefont{and} \bibinfo{author}{\bibfnamefont{J.~M.}
  \bibnamefont{{Cordes}}}, \bibinfo{journal}{\apj}
  \textbf{\bibinfo{volume}{596}}, \bibinfo{pages}{982} (\bibinfo{year}{2003}),
  \eprint{arXiv:astro-ph/0304365}.

\bibitem[{\citenamefont{{Weisberg} et~al.}(1986)\citenamefont{{Weisberg},
  {Armstrong}, {Backus}, {Cordes}, {Boriakoff}, and {Ferguson}}}]{wab+86}
\bibinfo{author}{\bibfnamefont{J.~M.} \bibnamefont{{Weisberg}}},
  \bibinfo{author}{\bibfnamefont{B.~K.} \bibnamefont{{Armstrong}}},
  \bibinfo{author}{\bibfnamefont{P.~R.} \bibnamefont{{Backus}}},
  \bibinfo{author}{\bibfnamefont{J.~M.} \bibnamefont{{Cordes}}},
  \bibinfo{author}{\bibfnamefont{V.}~\bibnamefont{{Boriakoff}}},
  \bibnamefont{and} \bibinfo{author}{\bibfnamefont{D.~C.}
  \bibnamefont{{Ferguson}}}, \bibinfo{journal}{\aj}
  \textbf{\bibinfo{volume}{92}}, \bibinfo{pages}{621} (\bibinfo{year}{1986}).

\bibitem[{\citenamefont{{Kedziora-Chudczer}
  et~al.}(2001)\citenamefont{{Kedziora-Chudczer}, {Jauncey}, {Wieringa},
  {Tzioumis}, and {Reynolds}}}]{kjw+01}
\bibinfo{author}{\bibfnamefont{L.~L.} \bibnamefont{{Kedziora-Chudczer}}},
  \bibinfo{author}{\bibfnamefont{D.~L.} \bibnamefont{{Jauncey}}},
  \bibinfo{author}{\bibfnamefont{M.~H.} \bibnamefont{{Wieringa}}},
  \bibinfo{author}{\bibfnamefont{A.~K.} \bibnamefont{{Tzioumis}}},
  \bibnamefont{and} \bibinfo{author}{\bibfnamefont{J.~E.}
  \bibnamefont{{Reynolds}}}, \bibinfo{journal}{\mnras}
  \textbf{\bibinfo{volume}{325}}, \bibinfo{pages}{1411} (\bibinfo{year}{2001}),
  \eprint{arXiv:astro-ph/0103506}.

\bibitem[{\citenamefont{{Spangler} and {Spitler}}(2004)}]{ss04}
\bibinfo{author}{\bibfnamefont{S.~R.} \bibnamefont{{Spangler}}}
  \bibnamefont{and} \bibinfo{author}{\bibfnamefont{L.~G.}
  \bibnamefont{{Spitler}}}, \bibinfo{journal}{Physics of Plasmas}
  \textbf{\bibinfo{volume}{11}}, \bibinfo{pages}{1969} (\bibinfo{year}{2004}).

\bibitem[{\citenamefont{{Lorimer} et~al.}(1995)\citenamefont{{Lorimer},
  {Yates}, {Lyne}, and {Gould}}}]{lylg95}
\bibinfo{author}{\bibfnamefont{D.~R.} \bibnamefont{{Lorimer}}},
  \bibinfo{author}{\bibfnamefont{J.~A.} \bibnamefont{{Yates}}},
  \bibinfo{author}{\bibfnamefont{A.~G.} \bibnamefont{{Lyne}}},
  \bibnamefont{and} \bibinfo{author}{\bibfnamefont{D.~M.}
  \bibnamefont{{Gould}}}, \bibinfo{journal}{\mnras}
  \textbf{\bibinfo{volume}{273}}, \bibinfo{pages}{411} (\bibinfo{year}{1995}).

\bibitem[{\citenamefont{{de Vos} et~al.}(2009)\citenamefont{{de Vos}, {Gunst},
  and {Nijboer}}}]{vgn09}
\bibinfo{author}{\bibfnamefont{M.}~\bibnamefont{{de Vos}}},
  \bibinfo{author}{\bibfnamefont{A.~W.} \bibnamefont{{Gunst}}},
  \bibnamefont{and}
  \bibinfo{author}{\bibfnamefont{R.}~\bibnamefont{{Nijboer}}},
  \bibinfo{journal}{IEEE Proceedings} \textbf{\bibinfo{volume}{97}},
  \bibinfo{pages}{1431} (\bibinfo{year}{2009}).

\bibitem[{\citenamefont{{Lonsdale} et~al.}(2009)\citenamefont{{Lonsdale},
  {Cappallo}, {Morales}, {Briggs}, {Benkevitch}, {Bowman}, {Bunton}, {Burns},
  {Corey}, {Desouza} et~al.}}]{lcm+09}
\bibinfo{author}{\bibfnamefont{C.~J.} \bibnamefont{{Lonsdale}}},
  \bibinfo{author}{\bibfnamefont{R.~J.} \bibnamefont{{Cappallo}}},
  \bibinfo{author}{\bibfnamefont{M.~F.} \bibnamefont{{Morales}}},
  \bibinfo{author}{\bibfnamefont{F.~H.} \bibnamefont{{Briggs}}},
  \bibinfo{author}{\bibfnamefont{L.}~\bibnamefont{{Benkevitch}}},
  \bibinfo{author}{\bibfnamefont{J.~D.} \bibnamefont{{Bowman}}},
  \bibinfo{author}{\bibfnamefont{J.~D.} \bibnamefont{{Bunton}}},
  \bibinfo{author}{\bibfnamefont{S.}~\bibnamefont{{Burns}}},
  \bibinfo{author}{\bibfnamefont{B.~E.} \bibnamefont{{Corey}}},
  \bibinfo{author}{\bibfnamefont{L.}~\bibnamefont{{Desouza}}},
  \bibnamefont{et~al.}, \bibinfo{journal}{IEEE Proceedings}
  \textbf{\bibinfo{volume}{97}}, \bibinfo{pages}{1497} (\bibinfo{year}{2009}),
  \eprint{0903.1828}.

\bibitem[{\citenamefont{{Cordes} and {Rickett}}(1998)}]{cr98}
\bibinfo{author}{\bibfnamefont{J.~M.} \bibnamefont{{Cordes}}} \bibnamefont{and}
  \bibinfo{author}{\bibfnamefont{B.~J.} \bibnamefont{{Rickett}}},
  \bibinfo{journal}{\apj} \textbf{\bibinfo{volume}{507}}, \bibinfo{pages}{846}
  (\bibinfo{year}{1998}).

\bibitem[{\citenamefont{{Cordes} et~al.}(2004)\citenamefont{{Cordes}, {Bhat},
  {Hankins}, {McLaughlin}, and {Kern}}}]{cbh+04}
\bibinfo{author}{\bibfnamefont{J.~M.} \bibnamefont{{Cordes}}},
  \bibinfo{author}{\bibfnamefont{N.~D.~R.} \bibnamefont{{Bhat}}},
  \bibinfo{author}{\bibfnamefont{T.~H.} \bibnamefont{{Hankins}}},
  \bibinfo{author}{\bibfnamefont{M.~A.} \bibnamefont{{McLaughlin}}},
  \bibnamefont{and} \bibinfo{author}{\bibfnamefont{J.}~\bibnamefont{{Kern}}},
  \bibinfo{journal}{\apj} \textbf{\bibinfo{volume}{612}}, \bibinfo{pages}{375}
  (\bibinfo{year}{2004}), \eprint{arXiv:astro-ph/0304495}.

\bibitem[{\citenamefont{{Rickett} et~al.}(1975)\citenamefont{{Rickett},
  {Hankins}, and {Cordes}}}]{rhc75}
\bibinfo{author}{\bibfnamefont{B.~J.} \bibnamefont{{Rickett}}},
  \bibinfo{author}{\bibfnamefont{T.~H.} \bibnamefont{{Hankins}}},
  \bibnamefont{and} \bibinfo{author}{\bibfnamefont{J.~M.}
  \bibnamefont{{Cordes}}}, \bibinfo{journal}{\apj}
  \textbf{\bibinfo{volume}{201}}, \bibinfo{pages}{425} (\bibinfo{year}{1975}).

\bibitem[{\citenamefont{{Cordes}}(1976)}]{c76}
\bibinfo{author}{\bibfnamefont{J.~M.} \bibnamefont{{Cordes}}},
  \bibinfo{journal}{\apj} \textbf{\bibinfo{volume}{210}}, \bibinfo{pages}{780}
  (\bibinfo{year}{1976}).

\bibitem[{\citenamefont{{Thompson} et~al.}(2011)\citenamefont{{Thompson},
  {Wagstaff}, {Brisken}, {Deller}, {Majid}, {Tingay}, and {Wayth}}}]{twb+11}
\bibinfo{author}{\bibfnamefont{D.~R.} \bibnamefont{{Thompson}}},
  \bibinfo{author}{\bibfnamefont{K.~L.} \bibnamefont{{Wagstaff}}},
  \bibinfo{author}{\bibfnamefont{W.~F.} \bibnamefont{{Brisken}}},
  \bibinfo{author}{\bibfnamefont{A.~T.} \bibnamefont{{Deller}}},
  \bibinfo{author}{\bibfnamefont{W.~A.} \bibnamefont{{Majid}}},
  \bibinfo{author}{\bibfnamefont{S.~J.} \bibnamefont{{Tingay}}},
  \bibnamefont{and} \bibinfo{author}{\bibfnamefont{R.~B.}
  \bibnamefont{{Wayth}}}, \bibinfo{journal}{\apj}
  \textbf{\bibinfo{volume}{735}}, \bibinfo{pages}{98} (\bibinfo{year}{2011}),
  \eprint{1104.4900}.

\bibitem[{\citenamefont{{Deneva} et~al.}(2009)\citenamefont{{Deneva}, {Cordes},
  {McLaughlin}, {Nice}, {Lorimer}, {Crawford}, {Bhat}, {Camilo}, {Champion},
  {Freire} et~al.}}]{dcm+09}
\bibinfo{author}{\bibfnamefont{J.~S.} \bibnamefont{{Deneva}}},
  \bibinfo{author}{\bibfnamefont{J.~M.} \bibnamefont{{Cordes}}},
  \bibinfo{author}{\bibfnamefont{M.~A.} \bibnamefont{{McLaughlin}}},
  \bibinfo{author}{\bibfnamefont{D.~J.} \bibnamefont{{Nice}}},
  \bibinfo{author}{\bibfnamefont{D.~R.} \bibnamefont{{Lorimer}}},
  \bibinfo{author}{\bibfnamefont{F.}~\bibnamefont{{Crawford}}},
  \bibinfo{author}{\bibfnamefont{N.~D.~R.} \bibnamefont{{Bhat}}},
  \bibinfo{author}{\bibfnamefont{F.}~\bibnamefont{{Camilo}}},
  \bibinfo{author}{\bibfnamefont{D.~J.} \bibnamefont{{Champion}}},
  \bibinfo{author}{\bibfnamefont{P.~C.~C.} \bibnamefont{{Freire}}},
  \bibnamefont{et~al.}, \bibinfo{journal}{\apj} \textbf{\bibinfo{volume}{703}},
  \bibinfo{pages}{2259} (\bibinfo{year}{2009}), \eprint{0811.2532}.

\bibitem[{\citenamefont{{Hankins} and {Rickett}}(1975)}]{hr75}
\bibinfo{author}{\bibfnamefont{T.~H.} \bibnamefont{{Hankins}}}
  \bibnamefont{and} \bibinfo{author}{\bibfnamefont{B.~J.}
  \bibnamefont{{Rickett}}}, in \emph{\bibinfo{booktitle}{Methods in
  Computational Physics. Volume 14 - Radio astronomy}}, edited by
  \bibinfo{editor}{\bibnamefont{{B.~Alder, S.~Fernbach, \& M.~Rotenberg}}}
  (\bibinfo{year}{1975}), vol.~\bibinfo{volume}{14}, pp.
  \bibinfo{pages}{55--129}.

\bibitem[{\citenamefont{{Johnston} et~al.}(2009)\citenamefont{{Johnston},
  {Feain}, and {Gupta}}}]{jfg09}
\bibinfo{author}{\bibfnamefont{S.}~\bibnamefont{{Johnston}}},
  \bibinfo{author}{\bibfnamefont{I.~J.} \bibnamefont{{Feain}}},
  \bibnamefont{and} \bibinfo{author}{\bibfnamefont{N.}~\bibnamefont{{Gupta}}},
  in \emph{\bibinfo{booktitle}{The Low-Frequency Radio Universe}}, edited by
  \bibinfo{editor}{\bibnamefont{{D.~J.~Saikia, D.~A.~Green, Y.~Gupta, \&
  T.~Venturi}}} (\bibinfo{year}{2009}), vol. \bibinfo{volume}{407} of
  \emph{\bibinfo{series}{Astronomical Society of the Pacific Conference
  Series}}, pp. \bibinfo{pages}{446--+}, \eprint{0903.4011}.

\bibitem[{\citenamefont{{Booth} et~al.}(2009)\citenamefont{{Booth}, {de Blok},
  {Jonas}, and {Fanaroff}}}]{bdj+09}
\bibinfo{author}{\bibfnamefont{R.~S.} \bibnamefont{{Booth}}},
  \bibinfo{author}{\bibfnamefont{W.~J.~G.} \bibnamefont{{de Blok}}},
  \bibinfo{author}{\bibfnamefont{J.~L.} \bibnamefont{{Jonas}}},
  \bibnamefont{and}
  \bibinfo{author}{\bibfnamefont{B.}~\bibnamefont{{Fanaroff}}},
  \bibinfo{journal}{ArXiv e-prints}  (\bibinfo{year}{2009}),
  \eprint{0910.2935}.

\bibitem[{\citenamefont{{Magro} et~al.}(2011)\citenamefont{{Magro},
  {Karastergiou}, {Salvini}, {Mort}, {Dulwich}, and {Zarb Adami}}}]{mks+11}
\bibinfo{author}{\bibfnamefont{A.}~\bibnamefont{{Magro}}},
  \bibinfo{author}{\bibfnamefont{A.}~\bibnamefont{{Karastergiou}}},
  \bibinfo{author}{\bibfnamefont{S.}~\bibnamefont{{Salvini}}},
  \bibinfo{author}{\bibfnamefont{B.}~\bibnamefont{{Mort}}},
  \bibinfo{author}{\bibfnamefont{F.}~\bibnamefont{{Dulwich}}},
  \bibnamefont{and} \bibinfo{author}{\bibfnamefont{K.}~\bibnamefont{{Zarb
  Adami}}}, \bibinfo{journal}{ArXiv e-prints}  (\bibinfo{year}{2011}),
  \eprint{1107.2516}.

\bibitem[{\citenamefont{{Deller}}(2010)}]{d10}
\bibinfo{author}{\bibfnamefont{A.}~\bibnamefont{{Deller}}}, in
  \emph{\bibinfo{booktitle}{Proceedings of the RFI Mitigation Workshop. 29-31
  March 2010. Groningen, the Netherlands}} (\bibinfo{year}{2010}), p.
  \bibinfo{pages}{PoS(RFI2010)035},
  \bibinfo{note}{http://pos.sissa.it/cgi-bin/reader/conf.cgi?confid=107},
  \eprint{1012.0325}.

\bibitem[{\citenamefont{{Wayth} et~al.}(2011)\citenamefont{{Wayth}, {Brisken},
  {Deller}, {Majid}, {Thompson}, {Tingay}, and {Wagstaff}}}]{wbd+11}
\bibinfo{author}{\bibfnamefont{R.~B.} \bibnamefont{{Wayth}}},
  \bibinfo{author}{\bibfnamefont{W.~F.} \bibnamefont{{Brisken}}},
  \bibinfo{author}{\bibfnamefont{A.~T.} \bibnamefont{{Deller}}},
  \bibinfo{author}{\bibfnamefont{W.~A.} \bibnamefont{{Majid}}},
  \bibinfo{author}{\bibfnamefont{D.~R.} \bibnamefont{{Thompson}}},
  \bibinfo{author}{\bibfnamefont{S.~J.} \bibnamefont{{Tingay}}},
  \bibnamefont{and} \bibinfo{author}{\bibfnamefont{K.~L.}
  \bibnamefont{{Wagstaff}}}, \bibinfo{journal}{\apj}
  \textbf{\bibinfo{volume}{735}}, \bibinfo{eid}{97} (\bibinfo{year}{2011}),
  \eprint{1104.4908}.

\bibitem[{\citenamefont{{Jackson} et~al.}(1989)\citenamefont{{Jackson},
  {Kundu}, and {White}}}]{jkw89}
\bibinfo{author}{\bibfnamefont{P.~D.} \bibnamefont{{Jackson}}},
  \bibinfo{author}{\bibfnamefont{M.~R.} \bibnamefont{{Kundu}}},
  \bibnamefont{and} \bibinfo{author}{\bibfnamefont{S.~M.}
  \bibnamefont{{White}}}, \bibinfo{journal}{\aap}
  \textbf{\bibinfo{volume}{210}}, \bibinfo{pages}{284} (\bibinfo{year}{1989}).

\bibitem[{\citenamefont{{Berger}}(2002)}]{b02}
\bibinfo{author}{\bibfnamefont{E.}~\bibnamefont{{Berger}}},
  \bibinfo{journal}{\apj} \textbf{\bibinfo{volume}{572}}, \bibinfo{pages}{503}
  (\bibinfo{year}{2002}), \eprint{arXiv:astro-ph/0111317}.

\bibitem[{\citenamefont{{Bastian} et~al.}(1990)\citenamefont{{Bastian},
  {Bookbinder}, {Dulk}, and {Davis}}}]{bbdd90}
\bibinfo{author}{\bibfnamefont{T.~S.} \bibnamefont{{Bastian}}},
  \bibinfo{author}{\bibfnamefont{J.}~\bibnamefont{{Bookbinder}}},
  \bibinfo{author}{\bibfnamefont{G.~A.} \bibnamefont{{Dulk}}},
  \bibnamefont{and} \bibinfo{author}{\bibfnamefont{M.}~\bibnamefont{{Davis}}},
  \bibinfo{journal}{\apj} \textbf{\bibinfo{volume}{353}}, \bibinfo{pages}{265}
  (\bibinfo{year}{1990}).

\bibitem[{\citenamefont{{Zarka}}(1998)}]{z98}
\bibinfo{author}{\bibfnamefont{P.}~\bibnamefont{{Zarka}}},
  \bibinfo{journal}{\jgr} \textbf{\bibinfo{volume}{103}},
  \bibinfo{pages}{20159} (\bibinfo{year}{1998}).

\bibitem[{\citenamefont{{Zarka} et~al.}(2004)\citenamefont{{Zarka}, {Farrell},
  {Kaiser}, {Blanc}, and {Kurth}}}]{zfk+04}
\bibinfo{author}{\bibfnamefont{P.}~\bibnamefont{{Zarka}}},
  \bibinfo{author}{\bibfnamefont{W.~M.} \bibnamefont{{Farrell}}},
  \bibinfo{author}{\bibfnamefont{M.~L.} \bibnamefont{{Kaiser}}},
  \bibinfo{author}{\bibfnamefont{E.}~\bibnamefont{{Blanc}}}, \bibnamefont{and}
  \bibinfo{author}{\bibfnamefont{W.~S.} \bibnamefont{{Kurth}}},
  \bibinfo{journal}{\planss} \textbf{\bibinfo{volume}{52}},
  \bibinfo{pages}{1435} (\bibinfo{year}{2004}).

\bibitem[{\citenamefont{{Ruf} et~al.}(2009)\citenamefont{{Ruf}, {Renno}, {Kok},
  {Bandelier}, {Sander}, {Gross}, {Skjerve}, and {Cantor}}}]{rrk+09}
\bibinfo{author}{\bibfnamefont{C.}~\bibnamefont{{Ruf}}},
  \bibinfo{author}{\bibfnamefont{N.~O.} \bibnamefont{{Renno}}},
  \bibinfo{author}{\bibfnamefont{J.~F.} \bibnamefont{{Kok}}},
  \bibinfo{author}{\bibfnamefont{E.}~\bibnamefont{{Bandelier}}},
  \bibinfo{author}{\bibfnamefont{M.~J.} \bibnamefont{{Sander}}},
  \bibinfo{author}{\bibfnamefont{S.}~\bibnamefont{{Gross}}},
  \bibinfo{author}{\bibfnamefont{L.}~\bibnamefont{{Skjerve}}},
  \bibnamefont{and} \bibinfo{author}{\bibfnamefont{B.}~\bibnamefont{{Cantor}}},
  \bibinfo{journal}{\grl} \textbf{\bibinfo{volume}{361}},
  \bibinfo{pages}{L13202} (\bibinfo{year}{2009}).

\bibitem[{\citenamefont{{Farrell} et~al.}(1999)\citenamefont{{Farrell},
  {Desch}, and {Zarka}}}]{fdz99}
\bibinfo{author}{\bibfnamefont{W.~M.} \bibnamefont{{Farrell}}},
  \bibinfo{author}{\bibfnamefont{M.~D.} \bibnamefont{{Desch}}},
  \bibnamefont{and} \bibinfo{author}{\bibfnamefont{P.}~\bibnamefont{{Zarka}}},
  \bibinfo{journal}{\jgr} \textbf{\bibinfo{volume}{104}},
  \bibinfo{pages}{14025} (\bibinfo{year}{1999}).

\bibitem[{\citenamefont{{Grie{\ss}meier}
  et~al.}(2007)\citenamefont{{Grie{\ss}meier}, {Zarka}, and {Spreeuw}}}]{gzs07}
\bibinfo{author}{\bibfnamefont{J.-M.} \bibnamefont{{Grie{\ss}meier}}},
  \bibinfo{author}{\bibfnamefont{P.}~\bibnamefont{{Zarka}}}, \bibnamefont{and}
  \bibinfo{author}{\bibfnamefont{H.}~\bibnamefont{{Spreeuw}}},
  \bibinfo{journal}{\aap} \textbf{\bibinfo{volume}{475}}, \bibinfo{pages}{359}
  (\bibinfo{year}{2007}), \eprint{0806.0327}.

\bibitem[{\citenamefont{{Hansen} and {Lyutikov}}(2001)}]{hl01}
\bibinfo{author}{\bibfnamefont{B.~M.~S.} \bibnamefont{{Hansen}}}
  \bibnamefont{and}
  \bibinfo{author}{\bibfnamefont{M.}~\bibnamefont{{Lyutikov}}},
  \bibinfo{journal}{\mnras} \textbf{\bibinfo{volume}{322}},
  \bibinfo{pages}{695} (\bibinfo{year}{2001}), \eprint{arXiv:astro-ph/0003218}.

\bibitem[{\citenamefont{{Pshirkov} and {Postnov}}(2010)}]{pp10}
\bibinfo{author}{\bibfnamefont{M.~S.} \bibnamefont{{Pshirkov}}}
  \bibnamefont{and} \bibinfo{author}{\bibfnamefont{K.~A.}
  \bibnamefont{{Postnov}}}, \bibinfo{journal}{\apss}
  \textbf{\bibinfo{volume}{330}}, \bibinfo{pages}{13} (\bibinfo{year}{2010}),
  \eprint{1004.5115}.

\bibitem[{\citenamefont{{Cordes} and {Lazio}}(1991)}]{cl91}
\bibinfo{author}{\bibfnamefont{J.~M.} \bibnamefont{{Cordes}}} \bibnamefont{and}
  \bibinfo{author}{\bibfnamefont{T.~J.} \bibnamefont{{Lazio}}},
  \bibinfo{journal}{\apj} \textbf{\bibinfo{volume}{376}}, \bibinfo{pages}{123}
  (\bibinfo{year}{1991}).

\bibitem[{\citenamefont{{Nita} et~al.}(2007)\citenamefont{{Nita}, {Gary},
  {Liu}, {Hurford}, and {White}}}]{ngl+07}
\bibinfo{author}{\bibfnamefont{G.~M.} \bibnamefont{{Nita}}},
  \bibinfo{author}{\bibfnamefont{D.~E.} \bibnamefont{{Gary}}},
  \bibinfo{author}{\bibfnamefont{Z.}~\bibnamefont{{Liu}}},
  \bibinfo{author}{\bibfnamefont{G.~J.} \bibnamefont{{Hurford}}},
  \bibnamefont{and} \bibinfo{author}{\bibfnamefont{S.~M.}
  \bibnamefont{{White}}}, \bibinfo{journal}{\pasp}
  \textbf{\bibinfo{volume}{119}}, \bibinfo{pages}{805} (\bibinfo{year}{2007}).

\bibitem[{\citenamefont{{Eatough} et~al.}(2010)\citenamefont{{Eatough},
  {Molkenthin}, {Kramer}, {Noutsos}, {Keith}, {Stappers}, and {Lyne}}}]{emk+10}
\bibinfo{author}{\bibfnamefont{R.~P.} \bibnamefont{{Eatough}}},
  \bibinfo{author}{\bibfnamefont{N.}~\bibnamefont{{Molkenthin}}},
  \bibinfo{author}{\bibfnamefont{M.}~\bibnamefont{{Kramer}}},
  \bibinfo{author}{\bibfnamefont{A.}~\bibnamefont{{Noutsos}}},
  \bibinfo{author}{\bibfnamefont{M.~J.} \bibnamefont{{Keith}}},
  \bibinfo{author}{\bibfnamefont{B.~W.} \bibnamefont{{Stappers}}},
  \bibnamefont{and} \bibinfo{author}{\bibfnamefont{A.~G.}
  \bibnamefont{{Lyne}}}, \bibinfo{journal}{\mnras}
  \textbf{\bibinfo{volume}{407}}, \bibinfo{pages}{2443} (\bibinfo{year}{2010}),
  \eprint{1005.5068}.

\end{thebibliography}
